\documentclass[fleqn,usenatbib]{mnras}

\usepackage{newtxtext,newtxmath}

\usepackage[T1]{fontenc}

\DeclareRobustCommand{\VAN}[3]{#2}
\let\VANthebibliography\thebibliography
\def\thebibliography{\DeclareRobustCommand{\VAN}[3]{##3}\VANthebibliography}

\usepackage{graphicx}	% Including figure files
\usepackage{amsmath}	% Advanced maths commands
\usepackage{amssymb}	% Extra maths symbols

\title[A census of OB stars within 1 kpc]{A census of OB stars within 1 kpc and the star formation and core collapse supernova rates of the Milky Way}

\author[A. L. Quintana, N.J. Wright and J. Mart\'inez Garc\'ia]{
Alexis L. Quintana$^{1,2}$\thanks{E-mail: alexis.quintana@ua.es}, Nicholas J. Wright$^{2}$ and Juan Mart\'inez Garc\'ia$^{2}$ \\
$^{1}$Departamento de Física Aplicada, Facultad de Ciencias, Universidad de Alicante, Carretera de San Vicente s/n, 03690 San Vicente
del Raspeig, Spain\\
$^{2}$Astrophysics Group, Keele University, Keele ST5 5BG, UK\\}

\date{Accepted 2025 January 10. Received 2024 December 10; in original form 2024 September 6}

\pubyear{2025}

\begin{document}
\label{firstpage}
\pagerange{\pageref{firstpage}--\pageref{lastpage}}
\maketitle

\begin{abstract}
OB stars are crucial for our understanding of Galactic structure, star formation, stellar feedback and multiplicity. In this paper we have compiled a census of all OB stars within 1 kpc of the Sun. We performed evolutionary and atmospheric model fits to observed spectral energy distributions (SEDs) compiled from astro-photometric survey data. We have characterized and mapped 24,706 O- and B-type stars ($T_{\rm eff} > 10,000$ K) within 1 kpc of the Sun, whose overdensities correspond to well-studied OB associations and massive star-forming regions such as Sco-Cen, Orion OB1, Vela OB2, Cepheus and Circinus. We have assessed the quality of our catalogue by comparing it with spectroscopic samples and similar catalogues of OB(A) stars, as well as catalogues of OB associations, star-forming regions and young open clusters. Finally, we have also exploited our list of OB stars to estimate their scale height (76 $\pm$ 1 pc), a local star formation rate of $2896^{+417}_{-1}$ M$_{\odot}$ Myr$^{-1}$ and a local core-collapse supernova rate of $\sim$15--30 per Myr. We extrapolate these rates to the entire Milky Way to derive a Galactic SFR of $0.67^{+0.09}_{-0.01}$ M$_{\odot}$ yr$^{-1}$ and a core-collapse supernova rate of 0.4--0.5 per century. These are slightly lower than previous estimates, which we attribute to improvements in our census of OB stars and changes to evolutionary models. We calculate a near-Earth core collapse supernova rate of $\sim$2.5 per Gyr that supports the view that nearby supernova explosions could have caused one or more of the recorded mass extinction events on Earth.
\end{abstract}


%%%%%%%%%%%%%%%%%%%%%%%%%%%%%%%%%%%%%%%%%%%%%%%%%%

%%%%%%%%%%%%%%%%% BODY OF PAPER %%%%%%%%%%%%%%%%%%

\section{Introduction}

OB stars are vital tools to study star formation (e.g. \citealt{Crowther2012}), stellar multiplicity (e.g. \citealt{Sana2012}) and the interstellar medium (ISM, e.g. \citealt{Hopkins2014}), as well as being important sources of stellar feedback \citep{Krumolhz2014,Dale2015}. In addition to generating H{\sc ii} regions and ionized bubbles (e.g. \citealt{DaleBonnell2010}), this feedback deposits considerable amounts of momentum and kinetic energy into the ISM \citep{KimOstriker2015}, and is thought to be responsible for slowing star formation by expelling molecular gas from star-forming regions \citep{Whitworth1979,Krumholz2019}. By ending their life in supernova explosions, massive OB stars ($\gtrsim$ 8--12 M$_{\odot}$ depending on metallicity, e.g. \citealt{Jones2013}) bring heavy elements into the ISM that are subsequently used for the formation of new stellar and planetary systems (e.g. \citealt{DeRossi2010}). 

Mapping out the distribution of OB stars is important for many reasons. Due to their short lifespan, OB stars remain close to their birth sites and therefore constitute valuable tracers of Galactic spiral arms (e.g. \citealt{Russeil2003,Quintana2023}). As the brightest members of OB associations (e.g. \citealt{Wright2020}), OB stars serve as key markers for identifying young stellar groups, offering valuable insights into their ages, spatial distributions, and reliable stellar kinematics.

Pioneering work to map OB stars across the Milky Way include the Case-Hamburg surveys (e.g. \citealt{Hardorp1959,Stephenson1971}), which were combined and updated by \citet{Reed2003} to produce a catalogue of 18,300 Galactic OB stars with \textit{UBV$\beta$} photometry (the Alma Luminous Stars catalogue, hereafter ALS I). Likewise, \citet{Goy1976} built a catalogue of 763 O-type stars with spectral types and associated H{\sc ii} regions, \citet{Garmany1982} built a similar catalogue of 765 O-type stars in the Milky Way, including cluster membership.

These early works however presented biases towards the visually brightest stars as they were compiled from spectroscopic observations. The arrival of \textit{Gaia} has led to improved censuses, as the combination of photometric, astrometric and spectroscopic data has enabled the production of much more extensive catalogues (e.g. \citealt{Chen2019,PantaleoniGonzalez2021,Zari2021}), including from the \textit{Gaia} consortium themselves in the form of the golden sample of OBA stars from \textit{Gaia} Data Release 3 (DR3, \citealt{GaiaDR3GoldenSample}). 

Improving the current census of OB stars is also important for identifying targets for spectroscopic follow-up, e.g., for William Herschel Telescope Enhanced Area Velocity Explorer (WEAVE, \citealt{WEAVE}) or the 4-metre Multi-Object Spectroscopic Telescope (4MOST, \citealt{4MOST}), as well as sources of gravitational wave (GW) emission, as future GW detectors are expected to be sensitive to events produced by core-collapse supernova (ccSN) explosions \citep{Radice2019,Srivastava2019,Powell2024}.
 
The aim of this work is to produce an unbiased census of OB stars within the local Milky Way, including within the solar neighbourhood (within 100 pc) where the visually brightest stars are lacking from many catalogues due to saturation. We have divided this paper as follows. In Section \ref{identification}, we outline the selection process of the candidate OB stars within 1 kpc and the method applied to characterize them, before evaluating the incompleteness of our catalogue and assessing the quality of our results through a comparison with spectroscopic catalogues. In Section \ref{analysis}, we analyse the broad features of our population of OB stars and compare it with other catalogues of OB(A) stars. In Section \ref{rates} we use our catalogue to estimate the star formation and ccSN rates within 1 kpc, discussing the implications of these results. Finally, in Section \ref{conclusions}, we summarise our findings.

\section{Identifying OB stars within 1 kpc}
\label{identification}

In this section we outline the method used to identify OB stars within 1 kpc of the Sun, in the plane of the Galactic disk\footnote{We chose to define our local volume as a cylinder arranged perpendicular to the Galactic disk rather than as a sphere to facilitate easier comparison with models and to reduce edge effects.}. In many historical studies, OB stars are defined as stars of spectral type B2 and earlier for dwarfs, B5 and earlier for giants, and of any O- and B-type for supergiants \citep{Morgan1951}, with this nomenclature followed in e.g. \citet{PantaleoniGonzalez2021}. In contrast, we define here an OB star as a star of spectral type of B9.5 or earlier, equivalent to an effective temperature greater than 10,000 K \citep{Mamajek}: our catalogue will thereby not strictly include young massive stars alone.

We perform this selection by first identifying a sample of candidate OB stars (Section \ref{data}), introduce the spectral energy distribution (SED) fitting code used to estimate their physical properties (Section \ref{SEDFitter}), and then present the general results of the process (Section \ref{genresults}), before assessing the quality of our catalogue of OB stars by quantifying its completeness (Section \ref{incompleteness}) and comparing our effective temperatures with spectroscopic temperatures (Section \ref{spectro}).

\subsection{Data and selection of candidate OB stars}
\label{data}

In this section we describe the selection of candidate OB stars within the area of interest, from the combination of \textit{Gaia} DR3 data and the \textit{Bright Star Catalogue} (hereafter referred as the BSC), which is used to compensate the saturation of \textit{Gaia} data at the brightest magnitudes.

\subsubsection{The \textit{Gaia} DR3 catalogue}
\label{maincatalogue}

To identify candidate OB stars stars for SED fitting we gathered astrometry and photometry from \textit{Gaia} DR3 \citep{GaiaDR3} and complemented this with additional data.

We started by selecting all \textit{Gaia} sources that fall within a distance of 1.1~kpc on the disk of the Galaxy, i.e., requiring that $\sqrt{X^2+Y^2} < 1.1$ kpc where $X = d \, \cos(l) \, \cos(b)$ and $Y = d \, \sin(l) \, \cos(b)$, and $d$ is the geometric line of sight distance from \citet{BailerEDR3}. We corrected the astrometry following the method from \citet{Maiz2021} and \citet{Maiz2022}. This allowed us to apply a more liberal cut on RUWE (re-normalised unit weight error) than is typically recommended, by only removing sources with RUWE $> 8$ (instead of the usual threshold of 1.4). In addition, we only kept sources with $\frac{\varpi}{\sigma_{\varpi}} > 2$, where $\varpi$ and $\sigma_{\varpi}$ stand respectively for the corrected \textit{Gaia} DR3 parallax and its uncertainty.

An A0V star of absolute magnitude $M_{Ks} = 0.949$ mag \citep{Mamajek} will have a maximum $K_s$ of $\sim$12~mag at 1 kpc, assuming a maximum $A_V$ of 10~mag (well above the maximum integrated line of sight extinction at 1~kpc across the vast majority of the sky), roughly translating into $A_{Ks} = 1$ mag. This value of $K_s$ is below the magnitude limit of 2MASS \citep{2MASS}, therefore almost every OB star in \textit{Gaia} DR3 should have a 2MASS counterpart.

For the \textit{Gaia} photometry we required $G_{\rm BP}$ and $G_{\rm RP}$ photometry to have $|C^*| < 3 \, \sigma_{C*}$ where $C^*$ is the corrected excess flux factor in the $G_{\rm BP}$ and $G_{\rm RP}$ bands and $\sigma_{C*}$ corresponds to the power-law on the $G$ band with a chosen 3$\sigma$ level \citep{Riello}. Furthermore, we corrected the \textit{Gaia} photometric uncertainties based on the calibration from \citet{Maiz2018}. Sources with $\sigma_G > 2 \, \sigma_{G_{\rm BP}} \sim 2 \, \sigma_{G_{\rm RP}}$ were considered as having non-valid $G$-band photometry as they correspond to partially unresolved binaries, and similarly for the $G_{\rm BP}$ and $G_{\rm RP}$ bands, thus having too high dispersion for a reliable \textit{Gaia} detection \citep{Maiz2023}. Finally we also applied the correction for photometric saturation from \citet{Riello}.

We cross-matched this sample with 2MASS\footnote{2 Micron All Sky Survey} \citep{2MASS}, IGAPS DR1\footnote{the INT Galactic Plane Survey} \citep{Drew,Mongui} and VPHAS+ DR2 \footnote{The VST Photometric H$\alpha$ Survey of the Southern Galactic Plane and Bulge} \citep{VPHAS} using a 1'' matching radius. We required that each individual 2MASS photometric band has a `good' photometric quality flag (i.e., `A', `B', `C' or `D'), excluded IGAPS photometry with associated classes that do not indicate a star or probable star \citep{Mongui}, and only used VPHAS+ photometry that was classified as `clean', i.e. with a significant detection and a good PSF fit \citep{VPHAS}\footnote{Throughout this paper we use the subscript `I' and `V' to distinguish the photometric bands from IGAPS and VPHAS+, respectively, and also use the subscripts `2M' for 2MASS photometry.}.

We used the near-IR photometry to perform an absolute magnitude cut to remove non OB stars, requiring that $M_{Ks} = K_s - 5 \, \log(d) + 5 < 1.95$ mag. This threshold equates to the magnitude of an A0V star \citep{Mamajek} at a distance of 1~kpc and with an extinction of 10~mag, as detailed above. If a source did not have good quality data in the $K_s$ band then we used the $H$ or $J$ bands, with appropriately adjusted thresholds. This cut on absolute magnitude allowed us to reduce the sample of $\sim$82 million sources to a manageable size of $\sim$3 million sources, making a more accurate cut based on absolute magnitude possible. We thus performed a \textit{Gaia} $G$ magnitude cut of:

\begin{equation}
\label{mgsec}
M_G = G - 5 \, \log(d) + 5 - 1.25 \, A_G < 1.5 \, \rm mag
\end{equation}

\noindent where the threshold of $M_G = 1.5$ mag is slightly fainter than that of an A0 star, $M_G = 1$~mag \citep{Mamajek}, to avoid being too conservative. The reddening, $A_G$, is derived from $A_V$ using the conversion factor of 0.843 from \citet{Zhang2023} and an additional factor of 1.25 to account for the temperature dependence of the $A_V$ to $A_G$ conversion \citep{Fouesnau2023}. $A_V$ is taken from the all-sky 3D extinction map of \citet{Edenhofer2024} using the geometric distances from \citet{BailerEDR3}\footnote{Some stars had a distance beyond the 1.25~kpc range of the extinction map, but these are all at high Galactic latitude where extinctions are generally low and we instead use the maximum extinction along the line of sight.}.

We then performed a colour cut of $(BP-RP)_0 < 0.5$, using a threshold that is more relaxed than the $BP-RP$ value of -0.037 for an A0V star to be liberal.

Finally, we required at least one valid blue photometric band for a source to be retained ($G$, $G_{\rm BP}$, $g_{\rm I}$ or $g_{\rm V}$). We also identified 430 sources with valid \textit{Gaia} DR3 astrometry that corresponded to duplicates, identified as separate sources in \textit{Gaia} DR3 but not in other photometric surveys. For these stars, we only used their \textit{Gaia} photometry and discarded other available photometry due to source blending.

This lead to a sample of 195,266 candidate OB stars. The cuts based on data quality discarded $\sim$5000 sources, hence about 2.5\% of the sample. We will discuss further the losses due to data quality in Section \ref{incompleteness}.

\subsubsection{The Bright Stars Catalogue}
\label{brightcatalogue}

Due to saturation at the bright end of the \textit{Gaia} catalogue, $\sim$20\% of stars with $G < 3$~mag are missing from \textit{Gaia} DR3 \citep{Fabricius2021}. To address this we use the Yale Catalogue of Bright Stars \citep[BSC,][]{Schlesinger1930}, the most recent release of which \citep{Hoffleit1991} contains 9095 stars, of which 1808 are of spectral type `O' or `B'\footnote{There is a more recent Bright Stars Catalogue, named USNO, that also includes bright stars with missing \textit{Gaia} DR3 astrometry \citep{Zacharias2022}. However, we still favour the Yale version because the USNO BSC improved the proper motions from Hipparcos but not its parallaxes (and we are only exploiting the latter in this work). Moreover, the USNO BSC includes 182 stars without \textit{Gaia} DR3 astrometry, without any information on their spectral type \citep{Zacharias2022}, while the Yale BSC contains 385 O- and B-type stars without valid \textit{Gaia} DR3 astrometry and is therefore more complete for our purpose.}. There are also five Wolf-Rayet (WR) stars in this catalogue, but only WR 11 is consistent with being closer than 1 kpc, so we only include this one \footnote{The scarcity of nearby WR stars can be explained by the fact that \textit{Gaia} struggles to infer their parallax. \citet{RateCrowther2020} included 12 WR stars within 1 kpc using \textit{Gaia} DR2 astrometry, but nearly all of them were flagged because of their astrometric data quality, and were not considered in the updated catalogue from \citet{Crowther2023} because of their bad \textit{Gaia} DR3 astrometry. Their absence will be accounted in Section \ref{incompleteness} when correcting the sample for incompleteness, notably because of bad astrometry.}. 

We cross-matched this sub-sample with \textit{Gaia} DR3 (using the \texttt{astroquery} package from Python and their identifier from the Henry Draper catalogue) and found 1683 out of 1809 stars in \textit{Gaia} DR3, including 1424 that were already part of our main catalogue of candidate OB stars (Section \ref{maincatalogue}). This means that there are 385 OB stars in the BSC that were missing from our initial sample, 259 of which were in \textit{Gaia} DR3 and 126 are not. 

For the stars that passed the criteria on data quality outlined in Section \ref{maincatalogue} but were rejected due to other reasons (e.g. they did not have a 2MASS counterpart) we included the star and used their \textit{Gaia} DR3 data.

For the 126 stars in the BSC that were not in \textit{Gaia} DR3, and the 174 stars that did not have valid \textit{Gaia} DR3 astrometry, we instead utilised astrometry from, in order of preference, \textit{Gaia} DR2 \citep{GaiaDR2}\footnote{For \textit{Gaia} DR2, we considered a source to have good astrometry if $RUWE < 1.4$, and corrected the observed parallax by the zero-point value of -0.029 mas from \citet{Lindegren2018}.}, \textit{Gaia} DR1\footnote{Correcting the observed parallax zero-point using \citet{ArenouGaiaDR1}.} or HIPPARCOS \footnote{For HIPPARCOS \citep{Hipparcos} data we used the improved version of the HIPPARCOS parallaxes from \citet{vanleuwen2007}, as well as their improved photometry if they were identified as part of a multiple system (provided that the quality of the solution was `A' or `B'), otherwise we used the photometry from the HIPPARCOS main catalogue.}.

For the stars in the BSC lacking valid photometry from \textit{Gaia} DR3, we instead use the photometry from, in order of preference, \textit{Gaia} DR2\footnote{For \textit{Gaia} DR2 we used the revised passbands from \citet{Maiz2018} for bright, $G_{\rm BP} > 10.87$ mag, and faint, $G_{\rm BP} < 10.87$ mag, sources. We also corrected the photometry for saturation based on \citet{Evans} if $G < $ 6 mag, otherwise we used the correction from \citet{Maiz2018}.}, \textit{Gaia} DR1\footnote{For \textit{Gaia} DR1 we used the pre-launch \textit{Gaia} bandpasses from \citet{Jordi2010}.} or HIPPARCOS\footnote{For HIPPARCOS we used the photometric bands $H_P$, $B_T$ and $V_T$ with the refined photometric passbands from \citet{Weiler2018}.}.

%We further complement this list by considering W-R stars, noting the scarcity of such objects in the BSC, but also from our main catalogue. \citet{RateCrowther2020} listed 12 W-R stars within 1 kpc with updated distances from \textit{Gaia} DR2, including WR 11. However, most of these stars were flagged because of their astrometry, and were not included in the updated census from \citet{Crowther2023} due to also having a warning flag on \textit{Gaia} DR3 astrometry as well. Investigating these stars one by one, we confirm that most of these WR stars are either too distant or have too unreliable astrometry to be included in our catalogue of OB stars within 1 kpc. Exceptions are WR 46-14 and WR 77p, with reliable \textit{Gaia} DR3 astrometry that did not pass our initial cylindrical cut in Section \ref{maincatalogue}, but whose lower error bars are consistent with being closer than 1 kpc, as well as WR 115 and WR 118-10, for which we favour \textit{Gaia} DR2 astrometry over \textit{Gaia} DR3 astrometry. All these four stars were thus added in our catalogue.

\begin{table}
	\centering
	\caption{Number of stars with valid data from the 385 O- and B-type stars from the BSC that are not part of the main catalogue of candidate OB stars. \label{BSCData}}
	\renewcommand{\arraystretch}{1.3} 
	\begin{tabular}{lcccccccr} 
		\hline
		Catalogue & Valid astrometry & Valid photometry  \\
		\hline
        \textit{Gaia} DR3 & 85 & 245 \\
        \textit{Gaia} DR2 & 73 & 26 \\
        \textit{Gaia} DR1 & 6 & 1 \\
        HIPPARCOS & 221 & 113 \\
		\hline
	\end{tabular}
\end{table}

Table \ref{BSCData} displays the breakdown of these sources from the BSC depending on the validity of their astrometry and photometry in each catalogue, thereby exploiting this data whenever good and available to characterize these stars.

The final catalogue of candidate OB stars was then compiled by combining the main catalogue described in Section \ref{maincatalogue} with the OB stars from the BSC, leading to 195,651 candidate OB stars.

\subsection{SED fitting}
\label{SEDFitter}

The SED fitting process compares the observed SED described in Section~\ref{data} with a forward-modelled SED that is computed using the combination of stellar evolutionary and atmospheric models to estimate stellar physical parameters. It is described in detail in \citet{Quintana} and \citet{Quintana2023} and summarised below.

The fitting process is implemented using a Markov Chain Monte Carlo (MCMC) process \citep[using the \textit{emcee} package from Python,][]{Emcee} to sample the posterior distribution and a maximum-likelihood test to assess the quality of the fit. The fitted stellar parameters are the star's initial mass ($\log(M/M_{\odot}$)), age (parameterised as Fr(Age), the fractional age) and distance. An additional parameter, $\ln(\rm f)$, is included as a scaling uncertainty that facilitates the convergence of $\chi^2$ \citep{Emcee,Casey}. The likelihood and priors on these model parameters are:

\begin{equation}
\label{priors}
\ln (P(\theta)) =
\begin{cases}
\log (\frac{1}{2 \, L^3} \, d^2 \, \exp{(\frac{-d}{L}})) & \text{if }
\begin{cases}
-1.0 \leq  \log(M/M_{\odot}) \leq 2.0  \\
0.0 \leq \rm Fr( Age)  \leq 1.0 \\
0.0 \leq d \leq 5000.0 \, \rm pc \\
-10.0 \leq \ln(f) \leq 1.0 \\
\end{cases} \\ 
- \infty  & \text{otherwise}
\end{cases}
\end{equation}

\noindent the prior on distance, originating from \citet{Bailer2015} includes a scale length $L$ whose value was chosen to be 1.35 kpc, in order to set an exponentially decreasing volume density of stars.

For each iteration of the MCMC simulation a model SED is constructed from the initial mass and age, using stellar evolutionary and atmospheric models. For the stellar evolutionary models we use the solar metallicity models from \citet{Ekstrom}. For the stellar atmosphere models we use a combination of the BT-NextGen models \citep{Asplund2009,Allard2012} for $T_{\rm eff} = 3000-5000$ K, the Kurucz models \citep{Coelho} for $T_{\rm eff} = 6000-20,000$ K and the Tubingen Tubingen Non Local Thermodynamical Equilibirum (NLTE) Model Atmosphere Package \citep{Werner1999,Rauch,Werner} for $T_{\rm eff} = 21,000-50,000$ K, all at solar metallicity and $\log g = 4$\footnote{While some evolved massive stars do have a lower surface gravity than this, our tests using BT-NextGen models at $\log g = 2$ showed only very minor differences in the predicted SEDs. The maximum additional uncertainty that results from our choice to ignore differences in surface gravity are $\log(T_{\rm eff})$ of $\sim$0.001 dex and $<$ 0.1 dex for $\log(L/L_{\odot})$.}. The model atmosphere was used to predict synthetic absolute photometry, which was then reddened and placed at the modelled distance to provide a forward-modelled SED. The extinction used was taken from the 3D extinction map of \citet{Edenhofer2024} at the distance modelled, combined with the extinction laws from \citet{Fitzpatrick2019}, using the mean galactic value of $R_V = 3.1$\footnote{$R_V$ has been shown to significantly deviate from this average value within regions of intense UV emission, such as H{\sc ii} regions and cavities filled with hot gas (e.g. the Local Bubble), sometimes reaching values higher than 4 \citep{ExtinctionMaizApellaniz2024}. Given that our catalogue is dominated by field late B-type stars, therefore decoupled from star-forming regions, we still adopt this unique value of $R_V = 3.1$ as a reasonable approximation for most of our stars. In the future, we plan to explore various values of $R_V$ in order to further improve the quality of the SED fits.}. Compared to \citet{Quintana} and \citet{Quintana2023}, we use different band-passes for the \textit{Gaia} DR3 $G$-band for faint ($G > 13$ mag) and bright ($G < 13$ mag) sources to derive our synthetic photometry. These new \textit{Gaia} DR3 sensitivity curves are the results of the recalibration of \textit{Gaia} photometry, similar to that performed for \textit{Gaia} DR2 by \citet{Maiz2018}, but for \textit{Gaia} DR3 \citep{Weilerprep}, with an example of this correction presented in \citet{MaizApellaniz2024}.

Systematic uncertainties were added to the measured photometric uncertainties for each band during the fitting process. These were taken from \citet{Riello} for \textit{Gaia} DR3, \citet{Evans} for \textit{Gaia} DR2, \citet{Mironov2002} for HIPPARCOS, \citet{BarentsenArt} and \citet{VPHAS} for IGAPS and VPHAS+, and \citet{Skrutsie} for 2MASS. 

For each fit the parameter space was sampled using 1000 walkers and 400 iterations (discarding the first half as a burn-in). Following this we checked for convergence, with the lack of convergence identified by either having $\ln(\rm f) > -4$ or by having a 90\% confidence interval on $\log(T_{\rm eff})$ greater than 0.5 dex. If convergence hadn't been achieved we performed an additional 1200 iterations (discarding the first 1000 as a burn-in), repeating this process up to three times if necessary. The best-fit values of initial mass, age and distance were taken as the 50th percentile of the posterior distribution, with the 16th and 84th percentiles providing the lower and upper 1$\sigma$ error bars, respectively. $\log(T_{\rm eff})$ and $\log(L/L_{\odot})$ were extracted in the same way as additional, indirect products of the fitting processes.

\subsection{General results}
\label{genresults}

We fitted the 195,651 (195,266 from the \textit{Gaia} DR3 main catalogue and 385 from the BSC) candidate OB stars using the SED fitter. This resulted in 30,569 objects that we fit as OB stars, defined here as stars hotter than an effective temperature of 10,000 K (based on their median SED-fitted value), using the threshold from \citet{Mamajek}. We applied a final cut on this catalogue by removing stars with $\sqrt{X^2+Y^2} < 1$ kpc, where this time the distance used to compute $X$ and $Y$ corresponds to the median SED-fitted distance, reducing the sample to 159,133 candidate OB stars and 24,706 SED-fitted OB stars. 

\begin{figure*}
    \centering
    \includegraphics[scale = 0.085]{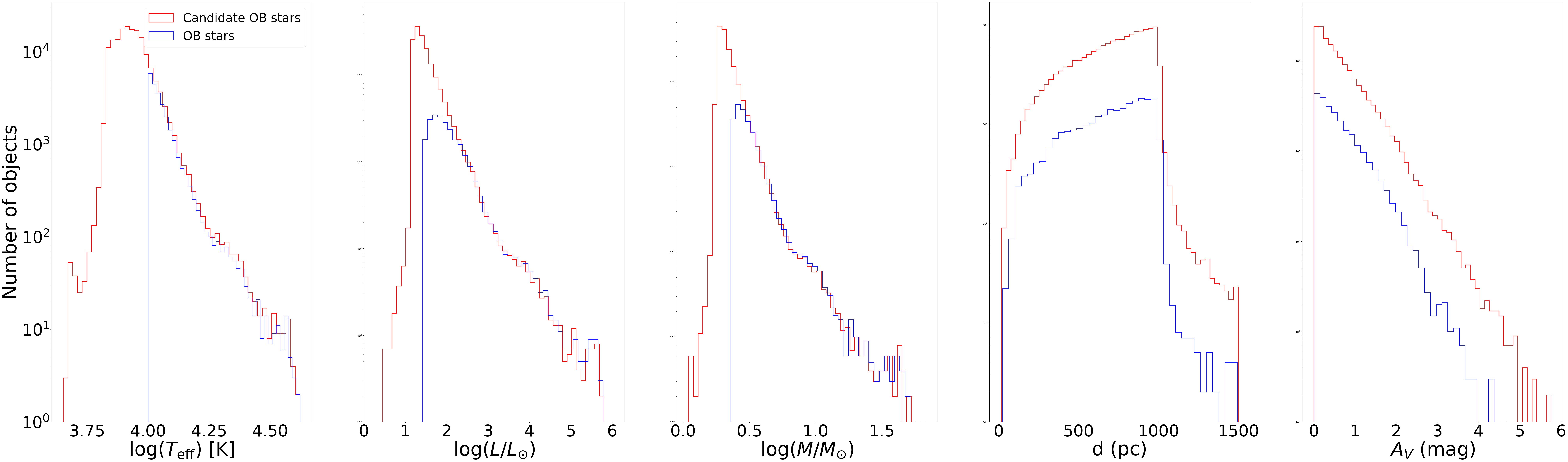}
    \caption{Histograms (in log scale) of the median SED-fitted parameters for the sample of 159,133 candidate OB stars (in red) and the 24,706 SED-fitted OB stars (in blue) within 1 kpc. \label{Fittedparam}}
\end{figure*}

The median values of the SED-fitted parameters for the 159,133 candidate OB stars within 1 kpc are displayed in Fig. \ref{Fittedparam}. Due to our photometric selection choices and the shape of the initial mass function this sample is dominated by A-type and late B-type stars. The median value of $\log(M / M_{\odot})$ is equal to 0.31 dex and the median value of $\log(L / L_{\odot})$ is equal to 1.40 dex. Most of the stars are located within 1 kpc, consistent with our selection choices, with some stars beyond the line-of-sight distance of 1 kpc because of the cylindrical nature of our selection volume. The extinction values tend to be small, with a median value at 0.45 mag and the majority lower than 3 mag. In Fig. \ref{Fittedparam} we have also displayed the 24,706 SED-fitted OB stars within 1 kpc to show their distribution

\subsection{Incompleteness}
\label{incompleteness}

To estimate the incompleteness of our catalogue we calculated the fraction of stars that were discarded during the selection process (e.g., due to data quality) as a function of magnitude. To calculate the fraction of \textit{Gaia} DR3 sources without parallaxes or proper motions (i.e., with 2-parameter solutions) we combined this approach with the table from the \textit{Gaia} archive (displaying this fraction for sources with $G \geq 9$ mag, \url{https://gea.esac.esa.int/archive/documentation/GDR3/Data\_processing/chap\_cu3ast/sec\_cu3ast\_quality/ssec\_cu3ast\_quality\_properties.html}), whilst using the $\sim$177,000 \textit{Gaia} DR3 sources with $G \leq 9$ mag to estimate the fraction for the faintest sources. Finally, thanks to the inclusion of the BSC, we assume a completeness of 100 \% for our stars with $G \leq 6.5$ mag.

\begin{figure}
    \centering
    \includegraphics[scale = 0.36]{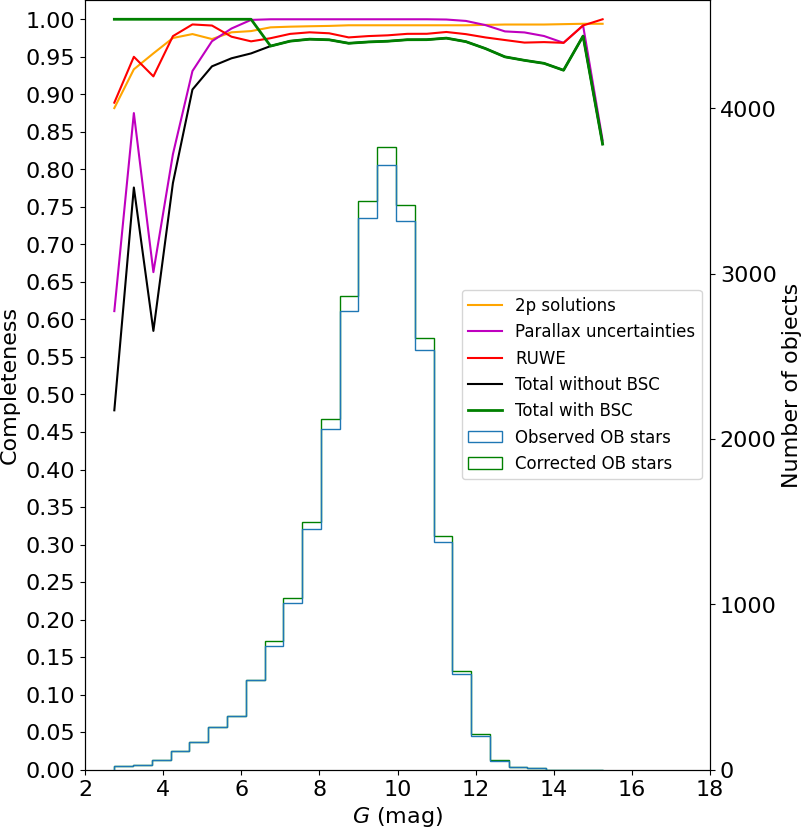}
    \caption{Observed (in blue) and completeness-corrected (in green) numbers of SED-fitted OB stars. within 1 kpc (right-hand Y axis). The lines at the top of the panel show the fraction of stars passing the individual selection steps based on data quality, as well as the total completeness that comes from the combination of all cuts, both with and without the inclusion of the BSC left-hand Y axis).}
    \label{IncompletenessSample}
\end{figure}

Our completeness fractions are shown as a function of $G$ magnitude in Figure \ref{IncompletenessSample}. The choice of more liberal cuts, compared with \citet{Quintana} and \citet{Quintana2023}, has allowed us to increase our completeness, typically reaching $>$95 \% across the magnitudes where most of the OB stars are encountered. 

We used these completeness estimates to correct the number of observed, SED-fitted OB stars as a function of $G$ magnitude in 0.5 mag bins.

\subsection{Comparison with spectroscopic temperatures}
\label{spectro}

To assess the quality of our SED fits we performed a comparison between our SED-fitted effective temperatures and spectroscopically-derived temperatures from the literature for our 159,133 candidate OB stars within 1 kpc. The samples used from the literature were as follows:

\begin{itemize}
    \item The Michigan catalogues of two-dimensional spectral types for HD stars \citep[][hereafter the `Houk catalogue']{Houk1975,Houk1978,Houk1982,Houk1988,Houk1999} was chosen for its size and homogeneity. We selected stars from this catalogue with a good measurement of spectral type (a value of 1 in spectral type quality), giving 74,275 stars, which were matched with 16,020 stars in our sample. Spectral types were converted to effective temperatures using the effective temperature scale of \citet{Martins} for O-type stars (observed scale), \citet{Trundle} for early B-type stars, \citet{Humphreys1984} for late B-type stars of luminosity classes `I' or `III' and from \citet{Mamajek} for all other later spectral types. We fixed an uncertainty of one spectral subclass, assumed a luminosity class of `V' when unspecified, used the spectral type of the primary star for binary systems, and finally interpolated between the luminosity classes `I', `III' and `V' for the luminosity classes of `II' and `IV'. 
    \item The Apache Point Observatory Galactic Evolution Experiment (APOGEE) DR17 spectroscopic sample \citep{Garcia,Abdu}, for which we selected stars with a measured $T_{\rm eff}$ and without any associated warnings, giving a sample of 9987 stars when cross-matched with our sample.
    \item The sample of stars with physical parameters from \textit{Gaia} DR3 \citep{Creevey2023}, ffrom the Extended Stellar Parametrizer for hot stars (ESP-HS) module, which derived stellar physical parameters with the BP/RP spectra with the Radial Velocity Spectrometer (RVS) spectra (if available), and the General Stellar Parametrizer from photometry (GSP-Phot) module (based on the low-resolution BP/RP spectra). Cross-matching this sample with ours, we obtained a list of 163,761 stars with a measurement in either the ESP-HS or GSP-Phot module in \textit{Gaia DR3}, thereby compiling a list of 107,940 and 75,629 stars for the two sub-samples, respectively. 
\end{itemize}

\begin{figure*}
    \centering
    \includegraphics[scale=0.2]{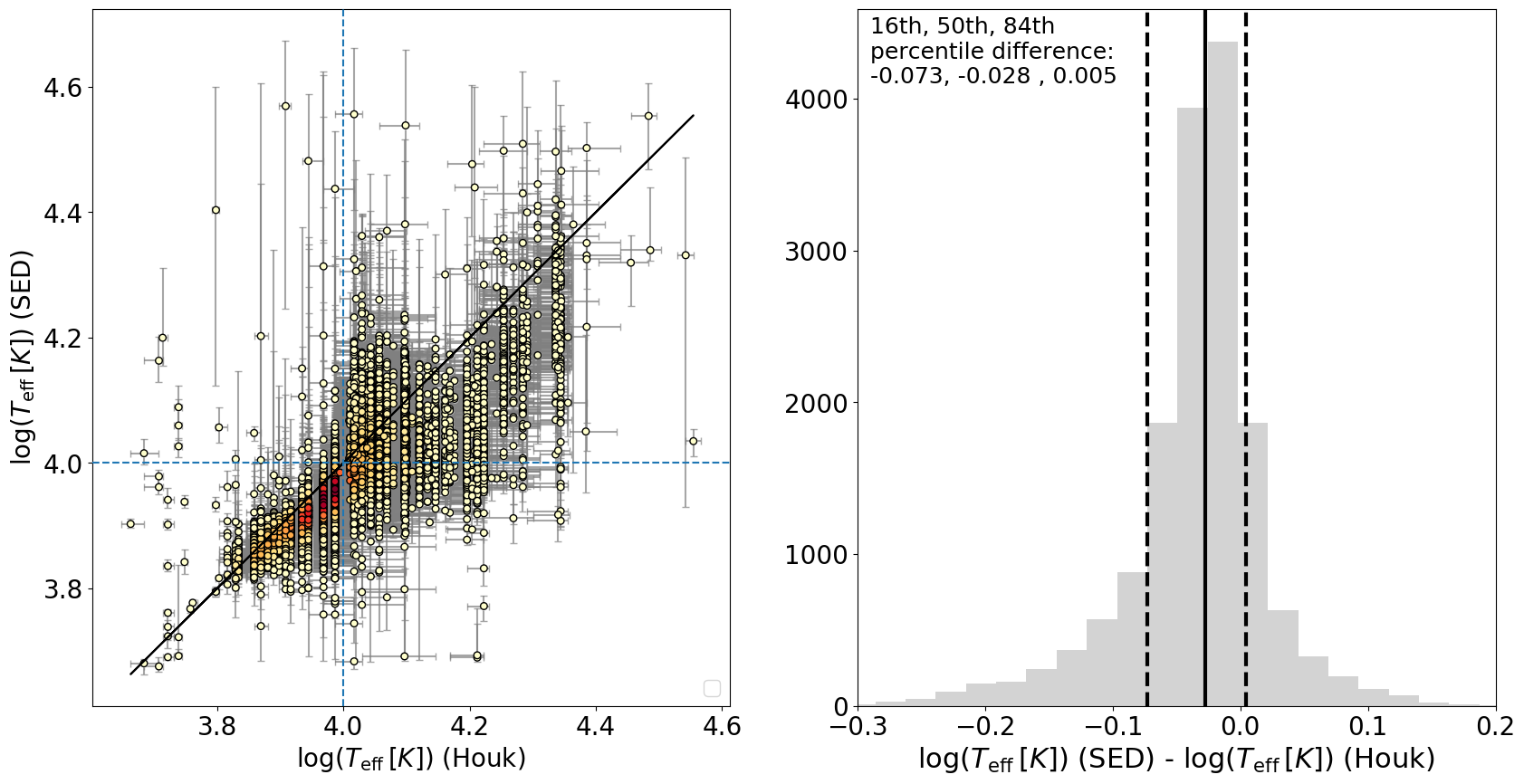}
     \includegraphics[scale=0.2]{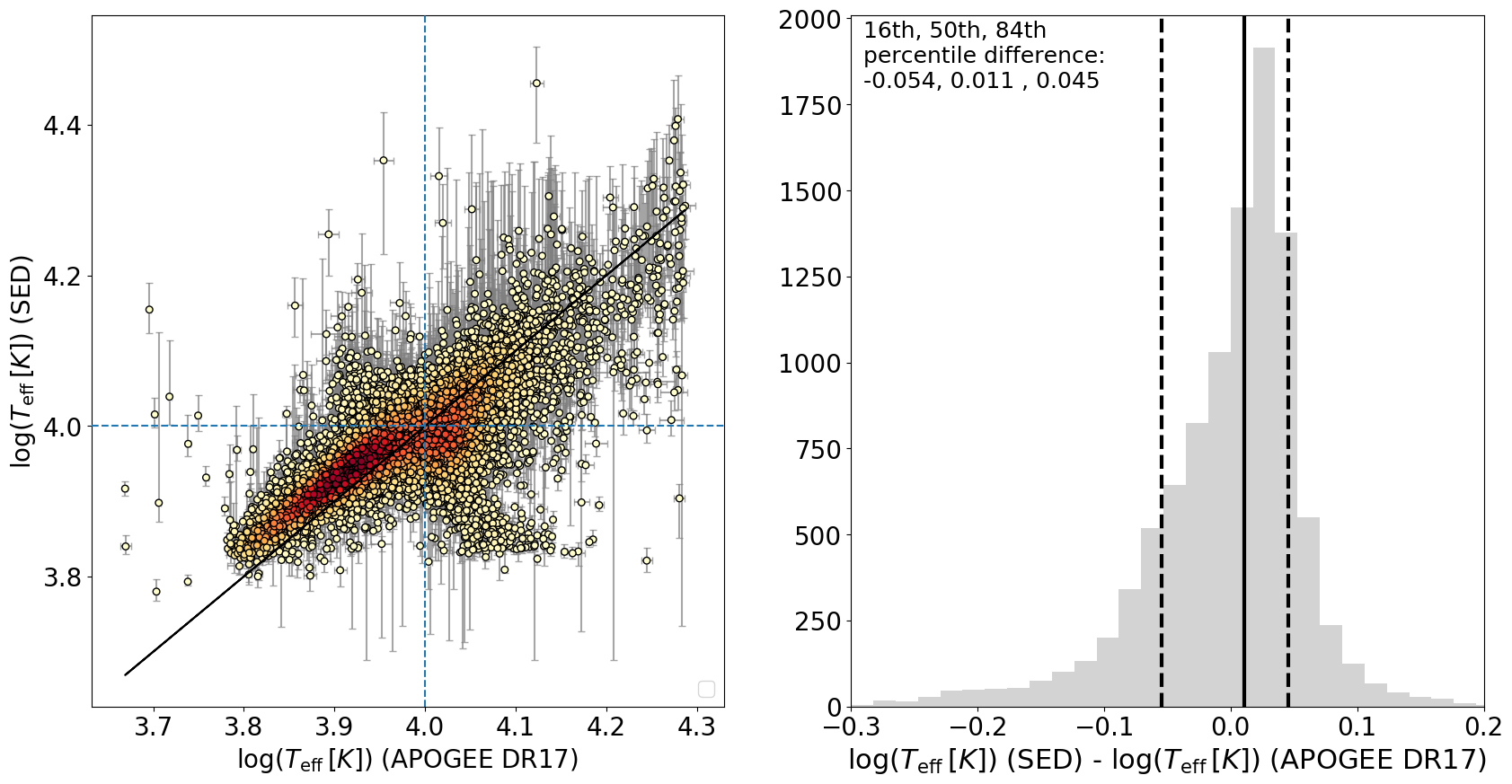}
     \includegraphics[scale=0.2]{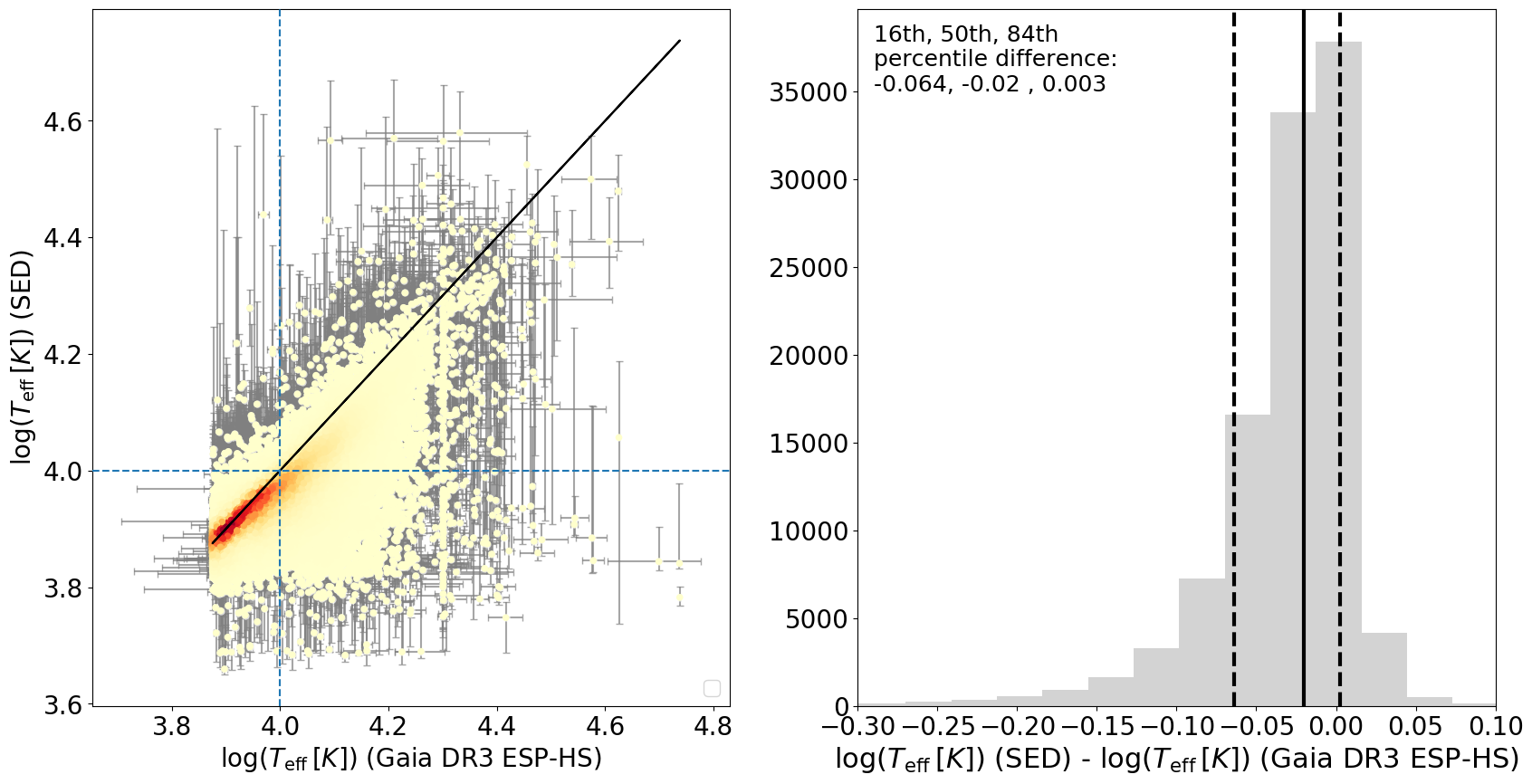}
     \includegraphics[scale=0.2]{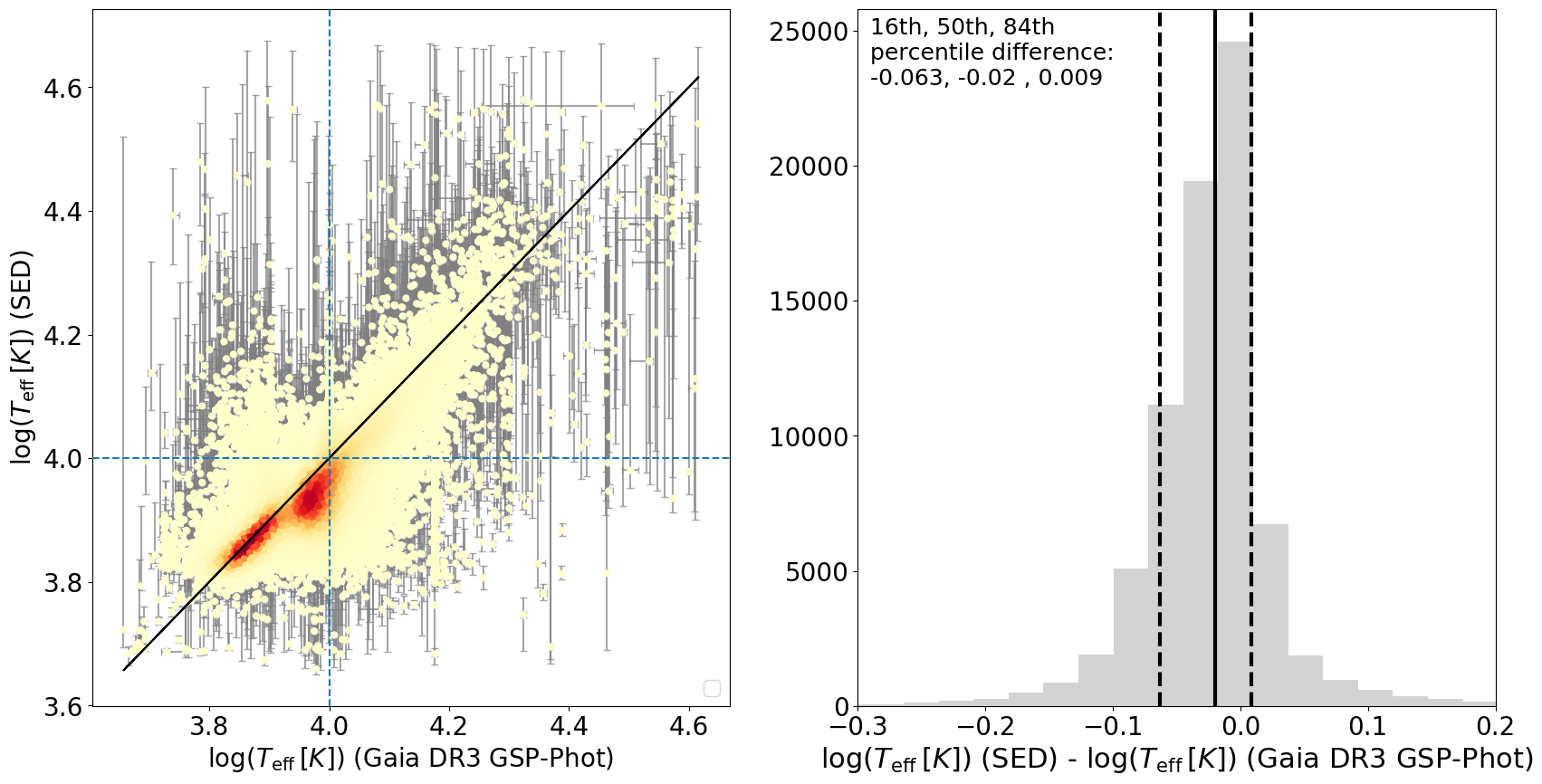}
    \caption{Far left and middle right panels: Comparison of our SED-fitted effective temperatures (ordinate) with the spectroscopic temperatures (abscissa) from the Houk (top left), APOGEE (top right), \textit{Gaia} DR3 ESP-HS  (bottom left) and GSP-Phot (bottom right) catalogues, all within 1 kpc. Each star has been colour-coded by its gaussian KDE on the graph. A black 1:1 line is shown as well as blue dashed lines corresponding to $\log(T_{\rm eff})$ = 4. Middle left and far right panels:  histograms of the difference between our median $\log(T_{\rm eff})$ and those of the corresponding spectroscopic sample, where we have indicated the 16th, 50th and 84th percentile differences on the top left.}
    \label{CompTeff}
\end{figure*}

The comparison between our SED-fitted temperatures and the spectroscopic temperatures from these samples is shown in Fig. \ref{CompTeff}. We see a generally good agreement, with median differences (biases) of the order of 0.01--0.03 dex (200--700 K at 10,000 K) and standard deviations of 0.03--0.05 dex (700--1200 K at 10,000 K).

There are however some minor disagreements evident in Figure \ref{CompTeff}. For instance, APOGEE does not fit effective temperatures above 20,000 K, which is visible in Fig. \ref{CompTeff} as an upper limit in temperature at that value, and means some stars around this limit may have underestimated temperatures. Apparent in Fig. \ref{CompTeff} in the comparison with APOGEE is the presence of `wings' extending perpendicular to the 1:1 line around $\log(T_{\rm eff})$ = 4. Investigation of the objects in these `wings' suggests they are due to poor spectral fits in APOGEE as we find that the same `wings' are evident in a comparison between APOGEE and \textit{Gaia} DR3 spectroscopic temperatures (see Appendix \ref{compgaiaapogee}).

The quality of the \textit{Gaia} DR3 Apsis modules has recently been discussed in \citet{Fremat2024}. They showed through a comparison with SIMBAD and spectroscopic surveys (e.g. LAMOST and the Gaia-ESO Survey) that stars classified as `hot' (OBA) in the Apsis modules can be contaminated by cooler stars such as white dwarfs, subdwarfs and RR Lyrae stars.

The Houk catalogues also have limitations. As explained in \citet{Skiff2014}, there are numerous cases in this catalogue where later-type stars have been wrongly classified as O- or B-type stars. Their tendency to overestimate temperatures thus gives weight to our method, since the upper left panel from Fig. \ref{CompTeff} shows that our SED-fitted temperatures tend to be lower than theirs, particularly for hot stars.

\section{Distribution of OB stars within 1 kpc}
\label{analysis}

In this section we use our sample to map out the distribution of OB stars within 1 kpc of the Sun, comparing it to previous samples of massive stars, as well as other tracers of the young stellar component of the Milky Way.

\subsection{Broad features of the population of OB stars}

\label{broadfeatures}
\begin{figure*}
    \centering
    \includegraphics[scale =0.5]{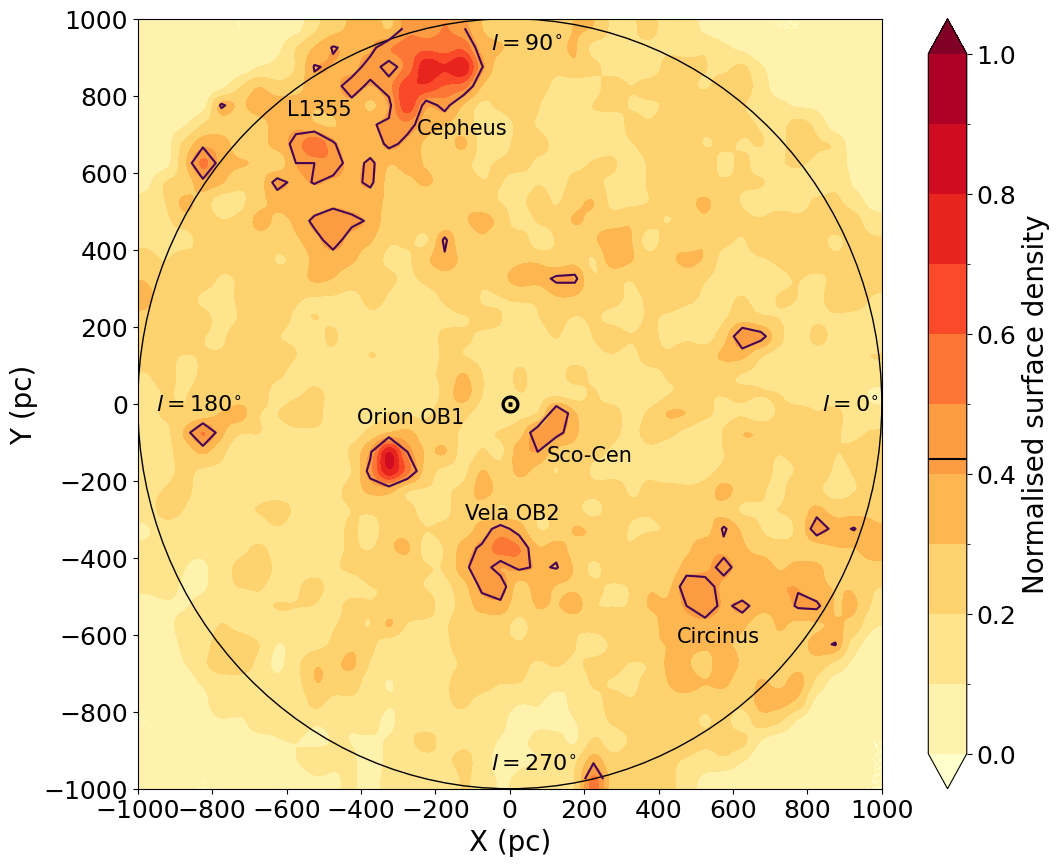}
    \caption{Surface density of the SED-fitted OB stars in Galactic Cartesian coordinates, smoothed with a gaussian interpolation, where the black circle delimits the 24,706 stars within 1 kpc. The colour scale has been normalised such that the maximum surface density value of 0.033 pc$^{-2}$ is equal to 1. The contours show over-densities that are likely to be real, with the selected threshold indicated on the colour bar. We have annotated the map with known star-forming regions and complexes.}
    \label{OBmap}
\end{figure*}

A normalised surface density map of our catalogue of SED-fitted OB stars is displayed in Fig. \ref{OBmap}. Given that the median uncertainty on the distance is $\sim$50 pc at the edges of the map (i.e. close to 1 kpc), we chose a bin size of 50 pc. The map shows a range of over-densities at different surface density levels, some of which are well known and others not.

To determine the level of over-density that is likely to be real (and below which we cannot distinguish between real over-densities and random fluctuations) we ran a simple Monte Carlo simulation. We took our sample of 24,706 OB stars and randomly assigned new positions for each star within 1~kpc (in any direction from their original positions), binned the stars on a 50~pc grid (to match with the bin size chosen above), and determined the highest surface density value. We performed this process 10,000 times and calculated the median value of the maximum surface density of 0.014 pc$^{-2}$, which corresponds to a normalised surface density of 0.421. We have chosen this value as the threshold above which the over-densities in Figure \ref{OBmap} likely represent real concentrations of OB stars and not random fluctuations, as it is improbable that any bin with a surface density larger than 0.014 pc$^{-2}$ can be attributed to random fluctuations. This threshold level is shown in the colour bar of Figure \ref{OBmap} and we have used it to draw contours and delineate the positions of the over-densities believed to be real in Fig. \ref{OBmap}.

Some of the most well-studied groups of young stars in the solar neighbourhood stand out clearly in this map, including the OB associations Sco-Cen, Vela OB2 and Orion OB1, the Cepheus complex \citep[which includes OB associations in Cepheus and Cassiopeia,][and appears to extend beyond the boundary of our study]{Wright2020}, an over-density of OB stars in the vicinity of Circinus \citep[which does not house any known OB associations at this distance, Circinus OB1 thought to be considerably more distant,][]{Wright2020} and another over-density that appears to be associated with the L1355 star forming region. In addition to these well-known regions we also identify a number of smaller over-densities.

\subsection{Comparison with other catalogues of OB stars}
\label{catalogues}

To further explore the completeness and quality of our catalogue, we compare it with four catalogues of OB stars from the recent literature. These catalogues are as follows:

\begin{itemize}
    \item \citet[][hereafter C19]{Chen2019} present a sample of O and early B-type stars photometrically-selected using VPHAS+ and \textit{Gaia} DR2 and complemented with a sample of spectroscopic O and early B-type stars from the literature. Their catalogue contains 1224 OB stars within 1 kpc of the Sun, of which 985 are found in our catalogue (80\%). About two thirds of the remaining stars are too faint to be main-sequence OB stars (they may be hot subdwarfs) while the remaining third of missing stars can be attributed to uncertainties in our $T_{\rm eff}$ estimations.
    \item \citet[][hereafter PG21]{PantaleoniGonzalez2021} updated the original Alma catalogue of OB stars (ALS I, \citealt{Reed2003}) with \textit{Gaia} DR2 astrometry and photometry to produce a new version of this catalogue, labelled ALS II. This catalogue includes 15,662 stars, from which we have selected the 2590 stars within 1 kpc that are either classified as `M' (likely massive stars) or `I' (high/intermediate-mass stars)\footnote{We included both subsets to ensure all sources with $T_{\rm eff} > 10,000$ K were included.}. A cross-match with our catalogue of SED-fitted OB stars gives 1312 stars in common. Most of our 23,393 remaining stars are too cool or faint to be included in PG21. Reversely, most of the stars that are only found in PG21 correspond to evolved (cool) massive stars, that we have not included in our catalogue.  
    \item \citet[][hereafter Z21]{Zari2021} identified OBA stars using \textit{Gaia} EDR3 and 2MASS photometry. They selected 24,342 OBA stars within 1 kpc of the Sun, of which 9345 are in our catalogue. Of the 14,997 not in our catalogue of OB stars, our SED fits reveal that the majority (13,251 or 88\%) are A-type stars. Conversely, there are 1558 OB stars in our catalogue not found in Z21, the majority of which are either late B-type stars that may have been missed by the selection method used by Z21 or are nearby ($d < 300$ pc) and missed by Z21 due to poor 2MASS photometry. 
    \item \citet[][hereafter G23]{GaiaDR3GoldenSample} used results from the GSP-Phot and ESP-HS Apsis modules from \textit{Gaia} DR3 \citep{Creevey2023} to build a sample of 3,023,388 OBA stars. They applied a threshold of $T_{\rm eff} >$ 7500 K using either the GSP-Phot or ESP-HS module temperatures, before filtering the sample to remove sub-luminous objects and halo contaminants. This sample is therefore cleaner than the simple \textit{Gaia} DR3 Apsis modules, as emphasized in \citet{Fremat2024}. G23 contains 303,678 OBA stars within 1 kpc of the Sun, 54,987 of which have $T_{\rm eff} > 10,000$ K using either the GSP-Phot or the ESP-HS temperature, but only 20,821 of these have matches in our catalogue. The majority of the remaining 34,166 are in our candidate OB star list, but were fitted with a lower temperature (though for many their uncertainties on $T_{\rm eff}$ extend above our temperature threshold). We cross-matched these lists with SIMBAD and found that 76\% of the 20,821 matching stars that were in SIMBAD had a spectral type of O or B, while only 38\% of the 34,166 non-matching stars in SIMBAD were OB stars. While this does suggest that our sample is missing some OB stars, the false-positive fraction of our sample is significantly smaller than for the G23. The differences between our catalogues are discussed in more details in Appendix \ref{compgaia}.
\end{itemize}

\begin{figure*}
    \centering
    \includegraphics[scale = 0.35]{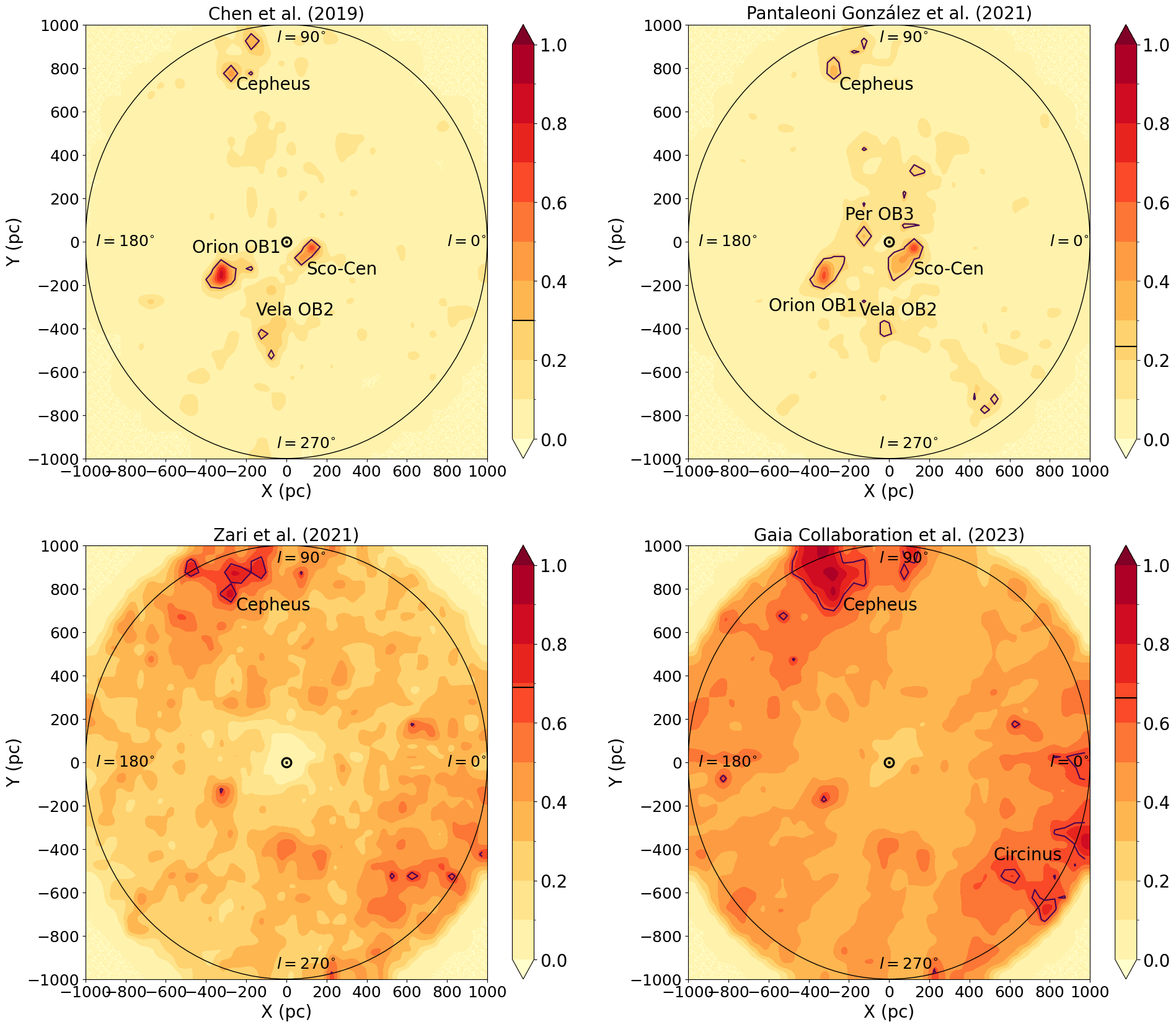}
    \caption{Normalised surface density map of the OB(A) stars within 1 kpc of the Sun for the OB star catalogues described in Section \ref{catalogues}. The maximum values respectively correspond to 0.008 pc$^{-2}$ for C19, 0.013 pc$^{-2}$ for PG21, 0.023 pc$^{-2}$ for Z21 and 0.226 pc$^{-2}$ for G23. We draw and annotate contours in the same manner as Figure \ref{OBmap}, indicating the threshold on the colour bar.}
    \label{MapCatalogues}
\end{figure*}

Figure \ref{MapCatalogues} shows the surface density distribution of the sources in the four comparison catalogues, whose colour scale we normalised similarly to Figure \ref{OBmap}, and where we have also drawn and annotated contours representing significant over-densities by the same method. There are many notable similarities and differences between the catalogues. The C19 and PG21 catalogues clearly show Orion OB1 and Sco-Cen, as well as Vela OB2 and Cepheus (although to a lesser extent compared with maps including later-type stars), and there is no evidence of an excess of OB stars in Circinus in C19 or PG21 as there is in our catalogue. This may be because Circinus lacks a significant number of O and early B-type stars or it could indicate a spatially-biased incompleteness in both the C19 and PG21 catalogues. Furthermore, PG21 is the only map where Per OB3 appears as a significant overdensity.

Z21 and G23, by contrast, are much noisier, only showing significant overdensities beyond the solar neighborhood. Sco-Cen is notably absent, most likely hinting at an incompleteness at the bright end of each catalogue, whilst Orion OB1 is slightly visible. Cepheus stands out in both maps, but more so in G23, where Circinus is more apparent as well. G23 also features an over-density at $(X,Y) \simeq (-800, -100)$ pc that we observe, though this is absent from the Z21 map.

%Both the Z21 and G23 catalogues show considerably more structure than our catalogue, C19 or PG21, but some of this may be noise. The G23 catalogue also shows some radial over-densities extending away from the Sun that are unlikely to be real and indicate spatially-biased contamination in the catalogue. Both Z21 and G23 clearly show Orion OB1, Vela OB2, the Circinus and Cepheus complexes, but in both catalogues Sco-Cen is notably absent, most likely hinting at an incompleteness at the bright end of each catalogue. Z21 and G23 both show an over-density at  $(X,Y) \simeq (+600, -200)$ pc, which we also recover, while G23 also features the over-density at $(X,Y) \simeq (-800, +100)$ pc that we observe, though this is absent from the Z21 map.

Notably, Z21 found a dearth of OB stars in the solar vicinity, which they attributed to either the presence of a Local Bubble, the absence of faint, late B-type stars in their sample, or the exclusion of bright nearby stars with poor 2MASS photometry (which their selection process required). We do not observe a low-density region around the Sun and it is only partly visible in the G23 map. Comparing these catalogues we find that Z21 are missing 424 stars in the solar vicinity ($d < 300$ pc) due to both poor 2MASS photometry and objects that are missing from the {\it Gaia} catalogue. Our catalogue does not rely on the availability of good 2MASS photometry and uses the BSC and {\it Hipparcos} data to replace the bright stars missing from {\it Gaia}, hence we do not suffer from this `local' incompleteness.

\subsection{Comparison with other catalogues of young stars}

Here we compare the distribution of OB stars in our catalogue with the distribution of young stars and groups of young stars found in other catalogues.

\subsubsection{OB associations}

\begin{figure*}
    \centering
    \includegraphics[scale =0.5]{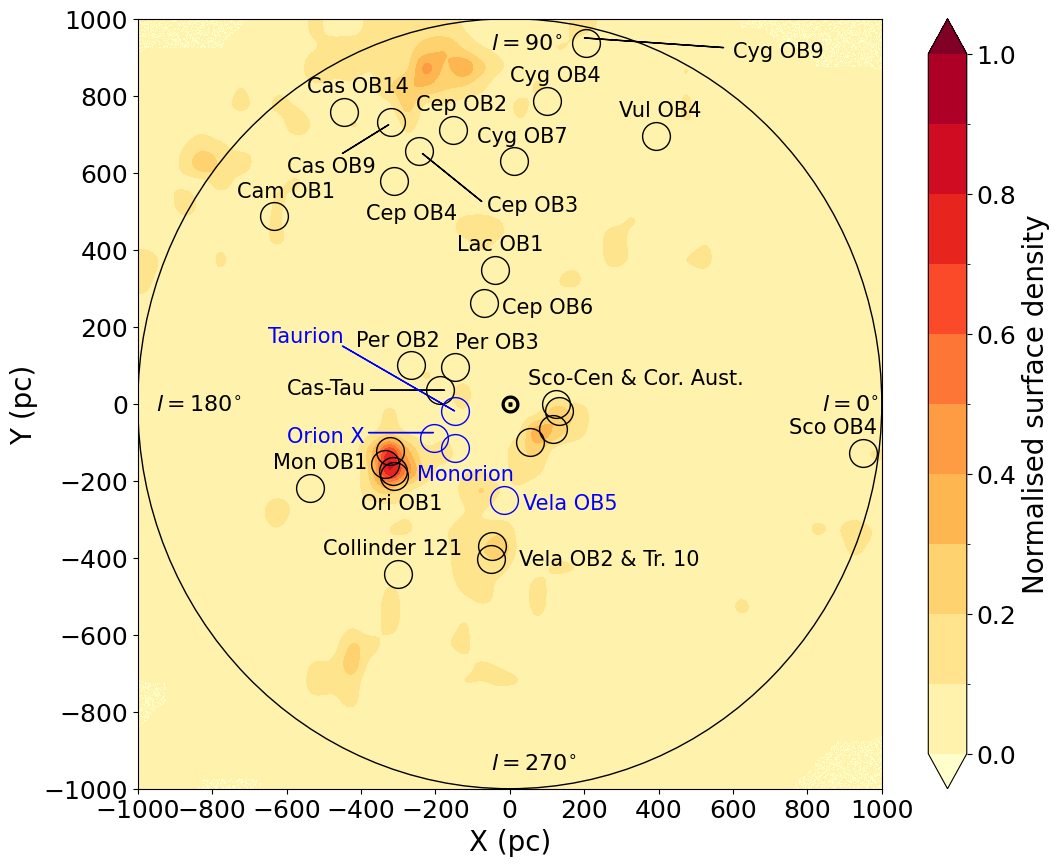}
    \caption{Cartesian coordinates of the historical OB associations listed in \citet{Wright2020} (black circles) and the OB groups from \citet{BouyAlves2015} (blue circles), plotted on the normalised surface density of the 3473 OB stars with a median SED-fitted age lower than 50 Myr.}
    \label{OBmapAssoc}
\end{figure*}

OB associations are kinematically-coherent groups of OB stars within which many OB stars are thought to form. \citet{Wright2020} lists 28 known OB associations within 1 kpc. The membership and position of these OB associations were established before the availability of \textit{Gaia} data, and since then some of the more distant systems have been shown not to be real \citep[e.g.,][]{Quintana,Chemel2022,Quintana2023,Fleming2023}, but for the present purposes we use the entire list. We complement this with the list of OB groups from \citet{BouyAlves2015}, specifically Monorion, Taurorion, Orion X (located in the foreground of Ori OB1) and Vela OB5 (in the foreground of Vela OB2).

The OB associations and OB groups are shown in Fig. \ref{OBmapAssoc} plotted on top of a surface density map of OB stars aged below 50 Myr from our catalogue (to provide a meaningful comparison, since most OB associations are young). This version of our OB star catalogue still shows the main stellar groups, but notably lacks any prominent overdensity in Circinus, suggesting this system may be older or be sufficiently low density to not stand out significantly in this map.

The majority of OB associations and all of the OB groups do not correspond to significant overdensities in our map. This is likely to either be because positions of the OB associations are incorrect (e.g., the historical positions of the Cepheus and Cassiopeia OB associations appear to be in the foreground of the main overdensity of OB stars in Cepheus in our map), are too small or too low-density to stand out on our map (e.g., Per OB2 and Orion X are relatively small), or are simply not real systems.

\subsubsection{Molecular clouds and star forming regions}

\begin{figure*}
    \centering
    \includegraphics[scale =0.5]{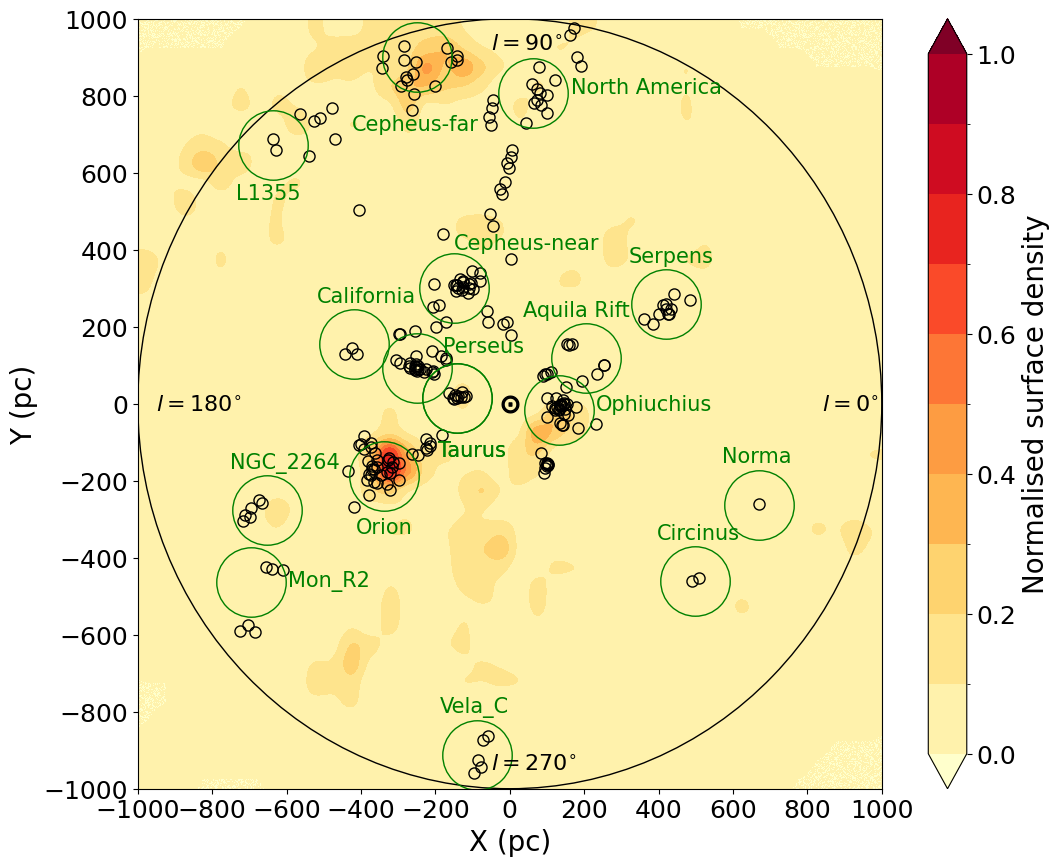}
    \caption{Cartesian coordinates of the molecular clouds listed in \citet{Zucker2020} (black circles), as well as the major SFRs of the area (green circles), plotted on the normalised surface density of the 3473 OB stars with a median SED-fitted age lower than 50 Myr.}
    \label{OBmapSFRs}
\end{figure*}

OB stars, particularly O-type stars, are very young and often found still in their birth star-forming region or in the vicinity of their parental molecular cloud. \citet{Zucker2020} provides a list of 241 molecular clouds in the solar neighbourhood, many of which are grouped together into larger star-forming regions. Figure \ref{OBmapSFRs} shows the distribution of these molecular clouds and star forming regions.

The picture revealed is similar to that of the OB associations, with some star-forming regions matching with clear over-densities of OB stars (e.g., the Orion and Cepheus-far star-forming regions) and others not (most likely because they are not massive enough to contain sufficient numbers of OB stars, e.g., Taurus). Notably, two of the OB star overdensities in Figure \ref{OBmap}, Circinus and L1355, coincide with known star-forming regions listed by \citet{Zucker2020}, supporting their validity.

\subsubsection{Open Clusters}

\begin{figure*}
    \centering
    \includegraphics[scale =0.45]{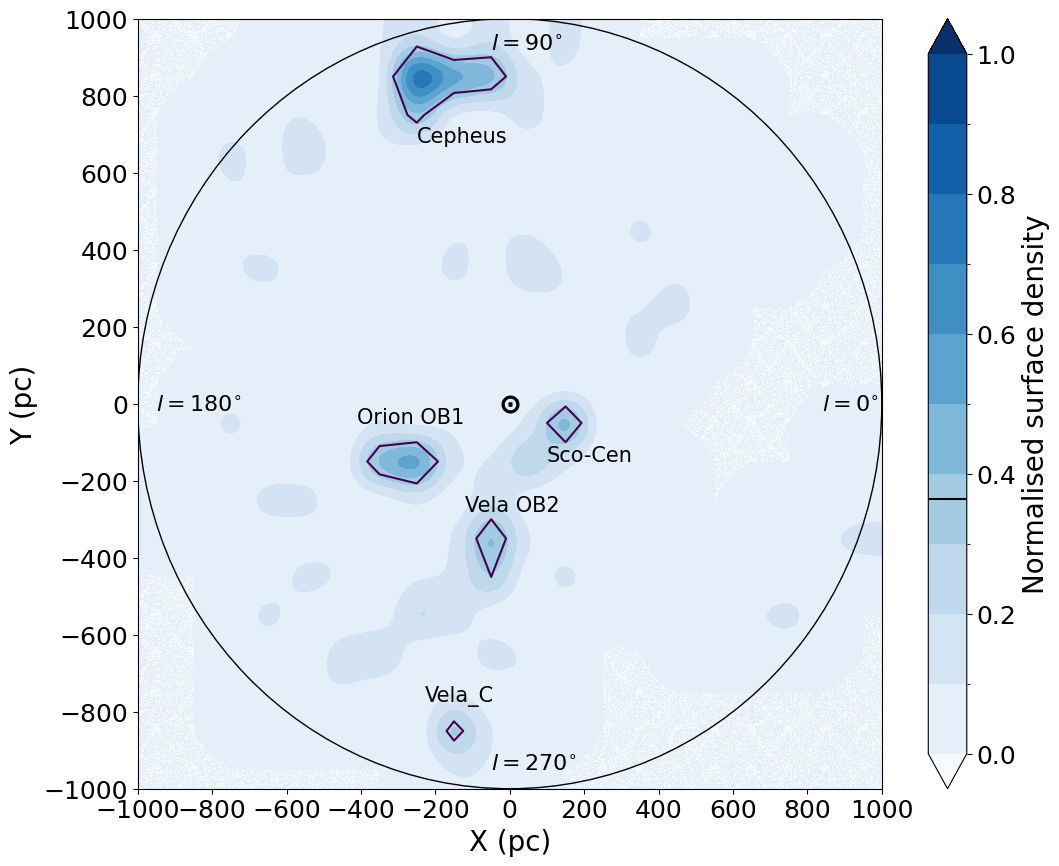}
    \caption{Normalised surface density map of the 191 reliable and young OCs from \citet{HuntReffert2024}, in Cartesian coordinates, with a bin size of 100 pc, as described in Section \ref{broadfeatures}. The maximum values respectively correspond to 0.0011 pc$^{-2}$. We drew and annotated contours in the same manner as Figure \ref{OBmap}, indicating the threshold on the colour bar.}
    \label{OBmapOCs}
\end{figure*}

Open clusters, which are often young, can contain many young OB stars. The catalogue from \citet{HuntReffert2023} is the most recent, complete and homogeneous sample of star clusters compiled using \textit{Gaia} data, including 1690 clusters within 1 kpc. We have exploited their updated catalogue from \citet{HuntReffert2024}, where bound OCs have been separated from unbound moving groups using their derived value of the Jacobi radius. From this subset of bound OCs, we have restricted the selection to those with a median isochronal age below 50 Myr (so they can serve as tracers of star formation), and followed the quality cuts from \citet[][i.e. a median CMD class greater than 0.5 as well as an astrometric S/N greater than 5$\sigma$]{HuntReffert2024}. This allowed us to produce a sample of 191 reliable and bound OCs within 1 kpc. We show the surface density map of bound, young OCs in Fig. \ref{OBmapOCs}. 

The picture unveiled shares some similarities with the surface density maps of the OB associations and SFRs, with Orion OB1, Vela OB2, Sco-Cen and Cepheus standing out again as over-densities. Compared with Figure \ref{OBmapSFRs}, Vela\_C is also noticeable, contrary to all the maps of OB(A) stars (Figures \ref{OBmap} and \ref{MapCatalogues}). 

\subsubsection{Young stars}

The SPYGLASS survey \citep{Kerr2021}, with the aim of mapping young stellar structures around the Sun, has a significant spatial overlap with our catalogue. This is particularly the case for the most recent version, SPYGLASS IV \citep{Kerr2023}, which includes 418,611 young stars ($<$ 50 Myr) within 1 kpc. \citet{Kerr2023} used a clustering algorithm to identify groups and clusters of young stars and divided their census between `clustered' and `extended' populations.

\begin{figure*}
    \centering
    \includegraphics[scale =0.45]{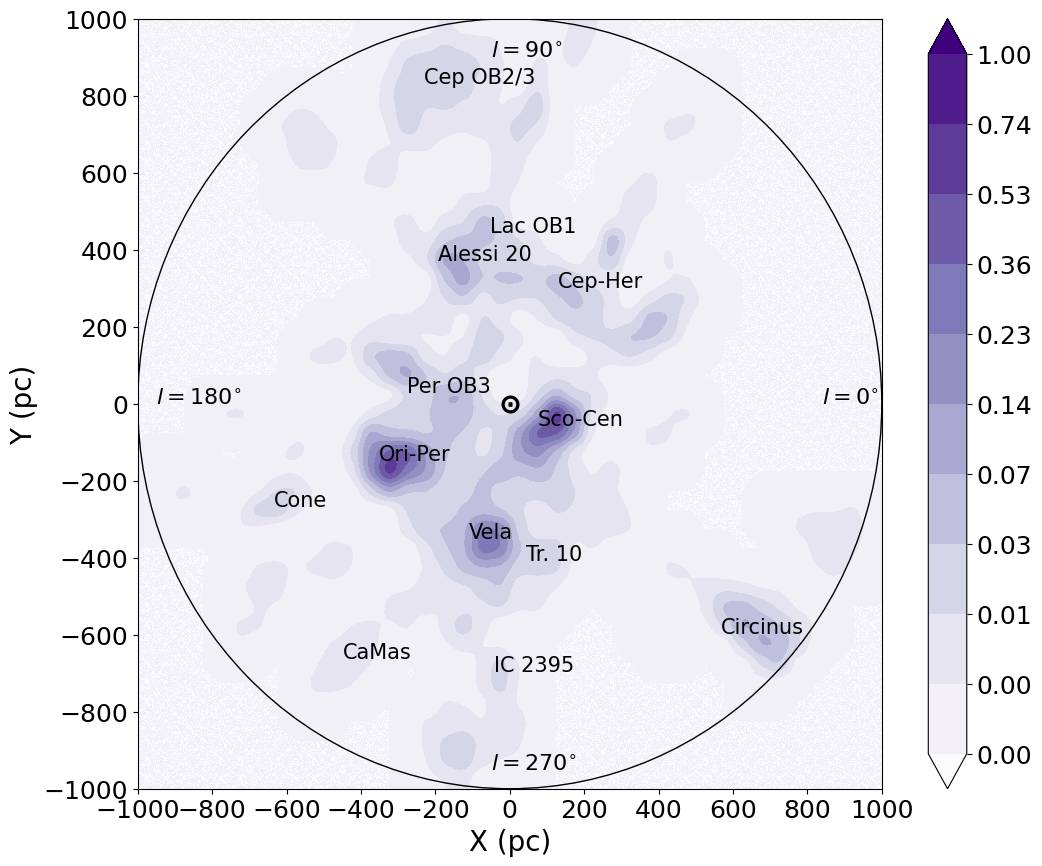}
    \caption{Normalised surface density map of the 36,182 young stars in the clustered population from SPYGLASS IV \citep{Kerr2023}, in Cartesian coordinates, with a bin size of 50 pc, as described in Section \ref{broadfeatures}. The colour scale was normalised by a power-law of index 0.35 to highlight the lower-density regions.}
    \label{OBmapYoung}
\end{figure*}

Figure \ref{OBmapYoung} shows the surface density distribution of the 36,182 young stars from SPYGLASS IV that belong to the clustered population, highlighting the main groups as in Fig. 8 from \citet{Kerr2023}. There are a number of overdensities that match with our OB star density map, including Sco-Cen, Vela and Orion, as well as Circinus to a lesser extent. On the other hand, there are some low-density groups missing from our map (Fig. \ref{OBmap}) such as the Cone Nebular Complex, Canis Major South (respectively Cone and CaMas in Fig. \ref{OBmapYoung}) and IC 2395, whilst their Cepheus region is of considerably lower density than ours. These differences are likely caused by these regions being less massive than other regions and therefore having fewer OB stars. On the other hand, the SPYGLASS clustered population does not show the Cepheus complex as prominently as our density map, which could be due to a higher incompleteness at the edge of the SPYGLASS survey area.

\section{Star formation and supernova rates in the Milky Way}
\label{rates}

In this section we exploit our catalogue of SED-fitted OB stars to calculate the star formation rate (SFR) and the core-collapse supernova rate in both the Solar Neighbourhood and the across the entire Milky Way. To do this we simulate a population of massive stars with a given star formation rate and compare it to our observed sample. We then extrapolate our results to the entire Galaxy and compare to literature results. 

\subsection{Simulated Population}
\label{simulations}

We started by generating a synthetic population of stars with an initial mass function from \citet[][sampled between 0.01 and 100 M$_{\odot}$ and with a high-mass exponent $\alpha = 2.3$]{IMFMasch} with ages randomly selected between 0 and $T_{max}$, the maximum age of a B9.5V star from \citet{Ekstrom}. We chose 380 Myr as the mean value between the model with no rotation (350 Myr) and with a rotation rate of $v_{\rm ini}/v_{\rm crit} = 0.4$ (410 Myr). Whether a star is a single star or a binary is determined randomly according to the mass-dependent binary fractions presented in \citet{DucheneKraus2013}. For the binary stars we randomly determine the mass of the secondary star from the mass ratio, $q = M_2/M_1$, assuming a flat distribution between 0 and 1. We do not model any binary interaction for simplicity.

We counted the total mass of all stars formed as well as the number of OB stars (those with $T_{\rm eff} > 10,000$ K at their age). If an OB star's age was above the maximum age of a star of that mass (of the primary star for multiple systems), we marked it as deceased, otherwise we retained it.

The minimum initial stellar mass for a star to undergo a ccSNe at the end of its life is $9.5^{+0.5}_{-1.2} M_{\odot}$ \citep{IbelingHeger2013}, though this is both metallicity and evolution dependent. Stars in our simulation with initial (primary) masses above this threshold were marked as having ended their lives in a SN. Whether there is an upper-mass limit for a star to explode in a supernova remains an ongoing debate \citep[e.g.,][]{Fryer1999,Heger2003}, though recent studies suggest an upper initial mass limit for a star to explode as a ccSN of between 20 and 40 M$_{\odot}$ \citep[e.g.,][]{Janka2012,DaviesBeasor2018,Schady2019}. We therefore estimate two ccSN rates, one without an upper-mass threshold for the progenitor (ccSN\_N rate) and one with an upper-mass threshold of 30 M$_{\odot}$ (ccSN\_U rate).

We scale the simulated number of living OB stars to match the observed, completeness-corrected (using the corrections calculated in Section \ref{incompleteness}) number of OB stars, and use the scaled total mass of all stars formed and the total time period of the simulation, $T_{max}$, to calculate the SFR. Using the galactic disk surface area of our studied area we then calculate the surface density formation rate, in units of M$_{\odot}$ Myr$^{-1}$ kpc$^{-2}$. The number of simulated stars that ended their lives as ccSN, per Myr, was also counted.

\begin{figure}
    \centering
    \includegraphics[scale = 0.38]{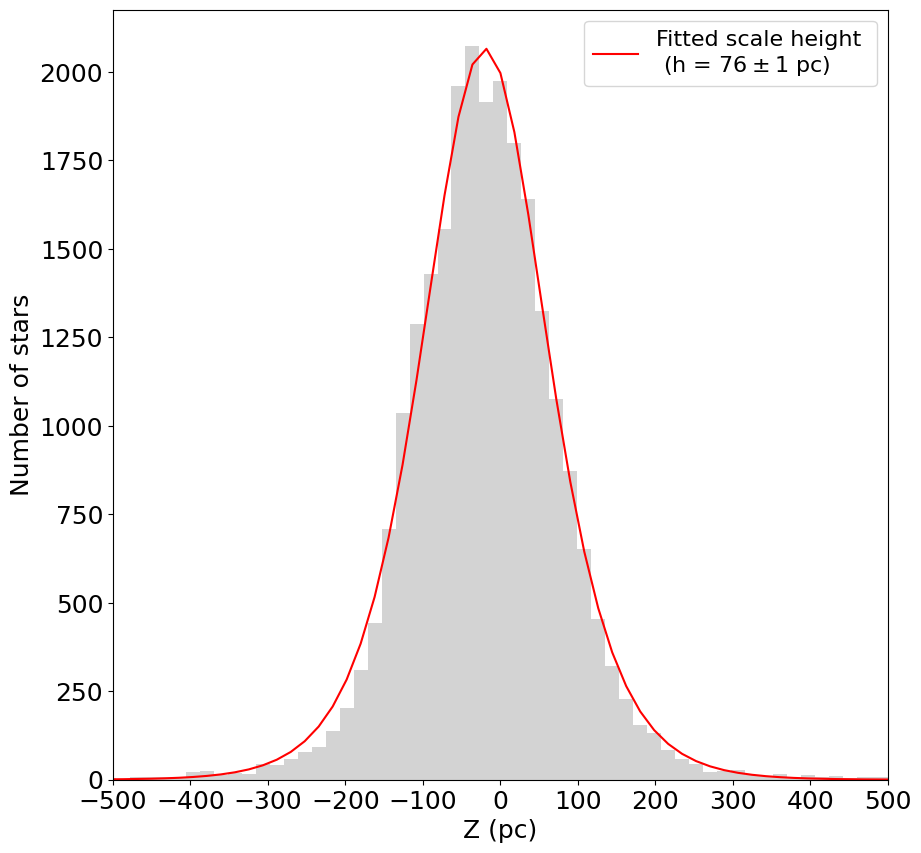}
    \caption{Sech$^2$ fit of the distribution of $Z$ values of the 24,706 OB stars.}
    \label{ScaleHeightOBStars}
\end{figure}

As an intermediate calculation, we calculate the scale height of OB stars using our sample. Fig. \ref{ScaleHeightOBStars} shows the $Z$ distribution of the OB stars in our sample, to which we have fitted a function of the following form (taken from \citealt{BobylevBajkova2016}):
\begin{equation}
\label{sechfunc}
\rho_z = \rho_0 \, sech^2 \left(\frac{Z-Z_{\odot}}{\sqrt{2} \, h}\right)
\end{equation}
\noindent using $Z_{\odot} = 20$ pc. 

With Eq. \eqref{sechfunc}, we find a scale height of $h = 76 \pm 1$ pc with uncertainties derived from a Monte Carlo experiment. Our scale height is larger than the 45~pc estimated by \citet{Reed2000}, but smaller than the value of $103.1 \pm 0.1$ pc calculated by \citet{KongZhu2008}.

To extrapolate our local SFR and ccSN rate to the entire Galaxy, we used equations 5 and 16 from \citet{Reed2005} to derive $\rho_{\rm plane}$, the present-day density of OB stars in the solar neighborhood, and the Galactic ccSN rate. To aid comparison we used the same values as \citet{Reed2005} for the distance of the Sun to the Galactic centre, $R_0$ (8.5 kpc, \citealt{DrimmelSpergel2001}), and used the aforementioned scale height $h$. \citet{Reed2005} assumed a two-component model for the Galaxy from \citet{DrimmelSpergel2001}, composed of a disk and a hole, with the exponential disk starting at a galactocentric radius $R_{\rm hole} = 0.5 \, R_0$, and with a radial density scale length of $H = 0.28 \, R_0$.  Finally, to estimate the ccSN rate within 20 pc of the Sun, that we label the {\it near-Earth rate} (discussed in Section \ref{massextinction}), we use the method from \citet{Fields2020}. All our results are displayed in Table \ref{TabRates} with our uncertainties.

%\begin{equation}
%R_{\rm Near-Earth} = R_{\rm Gal} \times \exp({\frac{-R_0}{H}}) \times \frac{d^3}{R_0^2 \, h}
%\end{equation}

%\noindent adopting $d = 20$ pc, and where $R_{\rm Near-Earth}$ and $R_{\rm gal}$ stand for the Near-Earth and the Galactic ccSN rate, respectively.

To estimate uncertainties for all these quantities we performed a Monte Carlo simulation with 1000 iterations, for each iteration of which we vary the following quantities:
\begin{enumerate}
    \item The observed number of OB stars in our sample, sampled from a Gaussian with a standard deviation equal to half the 16th-84th percentile range of the SED-fitted $\log(T_{\rm eff})$, and counting the number of stars with $\log(T_{\rm eff}) > 4$.
    \item The high-mass exponent of the IMF from \citet{IMFMasch}, varied according to its uncertainty, $\alpha = 2.3 \pm 0.1$.
    \item $T_{max}$, the upper age limit of the least-massive star in our OB sample, randomly selected between 350 and 410 Myr (chosen as the maximum age of a star of $T_{\rm eff} =$ 10,000 K in the evolutionary tracks from \citealt{Ekstrom}), depending on rotation rate.
    \item The upper initial mass limit for a star to end its life as a ccSN, which was randomly selected between 20 and 40 M$_{\odot}$.
\end{enumerate}

\subsection{Results and discussion}

\begin{table}
	\centering
	\caption{The star formation rate (SFR), surface density star formation rate (SFDR) and the core-collapse supernova (ccSN) rate derived from our catalogue of SED-fitted OB stars. The star formation and ccSN rates are presented for both the local 1-kpc volume (the {\it local} rate) and the entire Milky Way galaxy (the {\it Galactic} rate), whilst the ccSN rates are also presented in the local 20-pc volume (the {\it near-Earth} rate), and for both with (U) and without (N) an upper mass limit on the occurrence of a ccSN.
    \label{TabRates}}
	\renewcommand{\arraystretch}{1.3} 
	\begin{tabular}{lcccr} 
		\hline
		Parameter & Value & Units  \\
		\hline
        Local SFR & $2896^{+417}_{-1}$ & M$_{\odot}$ Myr$^{-1}$ \\
        Galactic SFR & $0.67^{+0.09}_{-0.01}$ & M$_{\odot}$ yr$^{-1}$ \\
        Local SFDR &  $922^{+133}_{-1}$ & M$_{\odot}$ Myr$^{-1}$ kpc$^{-2}$ \\
        Local ccSN\_U rate & ${16.5^{+6.7}_{-2.0}}$ & Myr$^{-1}$ \\
        Local ccSN\_N rate  & $20.6^{+8.1}_{-2.1}$ & Myr$^{-1}$ \\
       Galactic ccSN\_U rate  & $0.4 \pm 0.1$ & century$^{-1}$ \\
       Galactic ccSN\_N rate & $0.5 \pm 0.1$ & century$^{-1}$ \\
       Near-Earth ccSN\_U rate  & $2.0^{+0.9}_{-0.3}$ & Gyr$^{-1}$ \\
       Near-Earth ccSN\_N rate & $2.5^{+1.1}_{-0.3}$ & Gyr$^{-1}$ \\
		\hline
	\end{tabular}
\end{table}

Our results for the local SFR (and by extension, the local SFDR) are consistent with the estimate from \citet{BonattoBica2011}, who utilized a census of star clusters within 1 kpc to derive a SFR of $2500 \pm 500$ M$_{\odot}$ Myr$^{-1}$. Our local SFDR is approximately three times as large as the star formation density rate of 350 Myr$^{-1}$ kpc$^{-2}$ from \citet{LamersGieles2006} who used a census of star clusters within 600 pc. Recent, post-{\it Gaia} catalogues of star clusters (e.g., \citealt{HuntReffert2023}) have shown that older, pre-{\it Gaia} cluster catalogues were incomplete, implying that a SFR derived from such a catalogue would be underestimated. Our SFDR is consistent with, but at the low end, of the rate of 1000--3000 M$_{\odot}$ Myr$^{-1}$ kpc$^{-2}$ from \citet{LadaLada2003}, derived from embedded clusters within 500 pc, but considerably smaller than the value of 3000--7000 M$_{\odot}$ Myr$^{-1}$ kpc$^{-2}$ from \citet{MillerScalo1979} derived from field stars in the solar neighbourhood.

The value we have found for the Galactic SFR is equal to $0.67^{+0.09}_{-0.01}$ M$_{\odot}$ yr$^{-1}$. This result is consistent with the value of 0.68--1.45 M$_{\odot}$ yr$^{-1}$ from \citet{RobitaillWhithney2010}, but is lower than the values of 0.9--2.2 M$_{\odot}$ yr$^{-1}$ from \citet{MurrayRahman2010}, 1.9 $\pm$ 0.4 M$_{\odot}$ yr$^{-1}$ from \citet{ChomiukPovich2011}, 1.65 $\pm$ 0.19 M$_{\odot}$ yr$^{-1}$ from \citet{LicquiaNewman2015}, and $3.3^{+0.7}_{-0.6}$ M$_{\odot}$ yr$^{-1}$ from \citet{Zari2023}.

We estimate local ccSN rates of $16.5^{+6.7}_{-2.0}$ Myr$^{-1}$ (with an upper mass limit) and $20.6^{+8.1}_{-2.1}$ Myr$^{-1}$ (without an upper mass limit). These are in good agreement with the rate of $19.1^{+4.2}_{-5.4}$ Myr$^{-1}$ from \citet{Schmidt2014} that was derived from a sample of SN progenitors within 0.6 kpc\footnote{\citet{Schmidt2014} define SN progenitors as all stars of luminosity classes I and II, stars earlier than B9 for luminosity classes III and stars earlier than B4 for luminosity classes IV and V.}.

For the Galactic ccSN rate we estimate values of $0.4 \pm 0.1$ (with an upper mass limit) and $0.5 \pm 0.1$ (without an upper mass limit) per century. This value is notably lower than the value of 1--2 per century from \citet{Reed2005}, who based their estimate on a census of O and early B-type stars within 1.5 kpc. This difference is caused by three factors. Firstly, the census of O and early B-type stars compiled by \citet{Reed2005} is $\sim$40\% larger than ours (when the differences in spectral type range and the sample volume are accounted for). This difference is not surprising for a pre-{\it Gaia} sample where distances were based on an assumed absolute magnitude scale rather than parallaxes. Secondly, \citet{Reed2005} include stars in the spectral type range of O3--B2 in their sample, assuming the latter to have an initial mass of 10 $M_\odot$, while more recent models \citep{Ekstrom} give B2 stars an initial mass of $\sim$7 M$_\odot$. This difference means that \citet{Reed2005} would have included 45\% more stars as ccSN progenitors that we find are below the mass threshold from \citet{IbelingHeger2013}. Thirdly, \citet{Reed2005} used non-rotating evolutionary models to determine their stellar lifetimes (which were necessary for their ccSN rate calculations), while we used the rotating models from \citet{Ekstrom} that predict stellar lifetimes $\sim$25\% longer on average. The combination of these three effects explains the factor of 2.5 difference between our ccSN rate and that of \citet{Reed2005}.

Our ccSN rate is also lower than other estimates in the literature, though these are not based on the observed OB star population and are therefore less direct. For example, \citet{Cappellaro1993} estimated a ccSN rate of $1.37 \pm 0.74$ per century for the Milky Way, though this is extrapolated from studies of other galaxies. \citet{Adams2013} estimated a ccSN rate of $3.2^{+7.3}_{-2.6}$ per century using historical records of SNe, though this is highly uncertain. Finally, \citet{Rozwadowska2021} estimated a ccSN rate of $1.63 \pm 0.46$ per century using a combination of different methods, including extragalactic rates and historical data.

It is worth nothing that the Galactic ccSN rate we have calculated is highly dependent on the Galactic model used to extrapolate our local ccSN rate to the entire Galaxy \citep[as highlighted by][]{Reed2005}. For example, adopting $R_{\rm hole} = 2$ kpc instead of 4.25 kpc increases the Galactic ccSN rates by $\sim$ 50 \%. \citet{Reed2005} suggested that $R_{\rm hole} = 4.25$ kpc may be too large for OB stars, and if this was the case it would slightly increase our Galactic ccSN rate. Detecting the OB star populations towards the central regions of the Milky Way would help constrain the value of $R_{\rm hole}$. Visual extinction can be as large as $A_V \approx 30$ mag towards the Galactic centre (e.g. \citealt{Nogueras2019}), such that it is completely inaccessible to \textit{Gaia}. The proposed successor to \textit{Gaia}, GaiaNIR, would provide astrometry for $\sim$12 billion stars in the near-infrared, and would consequently be valuable to resolve the OB star population in the centre of the Milky Way, improving Galactic models \citep{HobbsHog2018,Hobbs2024}.

\subsection{Supernovae as triggers of mass extinction events on Earth}
\label{massextinction}

There is observational evidence that suggests supernova explosions have occurred within the solar neighbourhood within the astronomically recent past. \citet{MaizApellaniz2001} showed with HIPPARCOS proper motions that the position of Sco-Cen coincided with that of the Sun 5-7 Myr ago, and that this OB association has experienced $\sim$20 supernova explosions within the last 10-12 Myrs, generating enough feedback to produce the Local Bubble. An updated analysis with \textit{Gaia} data from \citet{Zucker2022} supports this scenario, with the supernovae starting to form the bubble at a slightly older time of $\sim$14 Myr ago. Measured outflows from Sco-Cen are also consistent with a relatively recent supernova explosion occurring in the solar neighbourhood \citep{Piecka2024}. Moreover, there is kinematic evidence of families of nearby clusters related to the Local Bubble, as well as with the GSH 238+00+09 supershell, suggesting that they produced over 200 supernovae within the last 30 Myrs \citep{Swiggum2024}.

Such SNe are thought to have influenced life on Earth. \citet{Benitez2002} linked the marine extinction at the Pliocene-Pleistocene boundary to a SNe from Sco-Cen 2-3 Myr ago, whilst \citet{Ertel2023} estimated that a similar event took place $\sim$7 Myr ago. Current evidence of these phenomena is supported by the amount of interstellar $^{60}$Fe detected in the Apollo lunar samples \citep{Fimiani2016}, the Antartic snow \citep{Koll2019} and the deep sea FeMn crust \citep{Wallner2021}.

While a SNe within a few hundred parsecs of Earth would have some minor effects on its atmosphere \citep[e.g.,][]{Firestone2014}, a much closer supernova within 20~pc would have a more catastrophic effect on Earth, destroying the Earth's ozone layer and leading to mass extinction \citep[e.g.,][]{Gehrels2003,Fields2020}.

There have been five mass extinction events on Earth over the last 500 Myr. Some of these extinction events have accepted causes, for example the Cretaceous–Paleogene extinction event, which occurred $66.043 \pm 0.043$ Myr ago \citep{Renne2013}, and led to the extinction of the dinosaurs, is commonly believed to have been caused by an asteroid impact in the Yucatan Peninsula (see \citealt{Morgan2022} for a recent review on this topic). Two extinction events have been specifically linked to periods of intense glaciation, potentially driven by dramatic reductions in the levels of atmospheric ozone that could have been caused by a near-Earth supernova \citep{Fields2020}, specifically the late Devonian and late Ordovician extinction events, 372 and 445 Myr ago, respectively \citep{BondGraspy2017}. Our near-Earth ccSN rate of $\sim$2.5 per Gyr is consistent with one or both of these extinction events being caused by a nearby SN.

% some do not have accepted causes (e.g., the Permian–Triassic and Triassic–Jurassic extinction events) 

\section{Conclusions}
\label{conclusions}

We have produced a new census of 24,706 O- and B-type stars ($>$ 10,000 K) within 1 kpc of the Sun. This was compiled using an improved version of our SED fitter, combining data from multiple photometric and astrometric surveys (including \textit{Gaia} DR3), with the Bright Stars Catalogue used to overcome {\it Gaia}'s incompleteness for bright stars. We estimate that our sample is $>$95\% complete, and from comparison with spectroscopic catalogues of OB stars we do not find any biases in estimating $T_{\rm eff}$.

Mapping the distribution of OB stars within 1 kpc of the Sun unveils stellar overdensities corresponding to well-known nearby OB associations and massive star-forming regions such as Sco-Cen, Orion OB1, Vela OB2, Cepheus and Circinus. These overdensities show a good level of overlap with known OB associations, star-forming regions and open clusters. In future work we will use this catalogue to identify coherent groups of young OB stars, such as OB associations, and catalogue their membership and properties.

Finally we exploited our list of OB stars to calculate the star formation and core-collapse supernova rates in the local Galaxy, and also extrapolated these values to the entire Milky Way. Our extrapolated Galactic SFR of $0.67^{+0.09}_{-0.01}$ M$_{\odot}$ yr$^{-1}$ is at the lower end of most previous estimates, which we attribute to our smaller (but more reliable) OB star catalogue. Our extrapolated ccSN rates of 0.4--0.5 per century are notably lower than most previous estimates, due to a combination of the smaller size of our OB catalogue and improved stellar evolutionary models. This reduced ccSN rate is important for modelling the rate of gravitational wave events from ccSN in the Milky Way \citep{Radice2019,Srivastava2019,Powell2024}. We also calculate a near-Earth ccSN rate (within 20 pc) of $\sim$2.5 Gyr$^{-1}$, which we argue is consistent with the rate of historical mass extinction events on Earth that are linked to ozone depletion and mass glaciation.

\section*{Acknowledgements}

We thank the anonymous referee for their detailed, insightful and valuable comments that have significantly improved this manuscript.

ALQ acknowledges the support from the Spanish Government Ministerio de Ciencia, Innovaci\'on y Universidades and Agencia Estatal de Investigación (MCIU/AEI/10.130 39/501 100 011 033/FEDER, UE) under grants PID2021-122397NB-C21/C22 and Severo Ochoa Programme 2020-2024 (CEX2019-000920-S). The research is also supported by MCIU with funding from the European Union NextGenerationEU and Generalitat Valenciana in the call Programa de Planes Complementarios de I+D+i (PRTR 2022), project HIAMAS, reference ASFAE/2022/017, as well as receipt of an STFC postgraduate studentship.

The authors would like to thank Jesús Maíz Apellániz for his recommendations about how to calibrate and select \textit{Gaia} data, and also for providing the corrected \textit{Gaia} DR3 photometric sensitivity curves. The authors would like to thank Ronan Kerr as well, for providing us his catalogue of his clustered population of young stars from SPYGLASS IV.

This paper benefited from the data processed by the Gaia Data Processing and Analysis Consortium (DPAC, https://www.cosmos.esa.int/web/gaia/dpac/consortium) and obtained by the Gaia mission from the European Space Agency (ESA) (https://www.cosmos.esa.int/gaia), as well as the INT Galactic Plane Survey (IGAPS) from the Isaac Newton Telescope (INT) operated in the Spanish Observatorio del Roque de los Muchachos, the observations made with ESO Telescopes at the La Silla Paranal Observatory under programme ID 177.D-3023, as part of the VST Photometric H$\alpha$ Survey of the Southern Galactic Plane and Bulge (VPHAS+, www.vphas.eu), alongside the Two Micron All Star Survey, which is a combined mission of the Infrared Processing and Analysis Center/California Institute of Technology and the University of Massachusetts.

Finally, this work also benefited from the use of \textit{TOPCAT} \citep{Topcat}, Astropy \citep{Astropy} and the Vizier and SIMBAD database, both operated at CDS, Strasbourg, France.

%%%%%%%%%%%%%%%%%%%%%%%%%%%%%%%%%%%%%%%%%%%%%%%%%%

\section*{Data Availability}
The data underlying this article will be uploaded to Vizier.



\bibliographystyle{mnras}
\bibliography{Bibliography} % if your bibtex file is called example.bib

%\begin{thebibliography}{99}
%\end{thebibliography}

%%%%%%%%%%%%%%%%%%%%%%%%%%%%%%%%%%%%%%%%%%%%%%%%%%

%%%%%%%%%%%%%%%%% APPENDICES %%%%%%%%%%%%%%%%%%%%%

\appendix

\section{Comparison between APOGEE and \textit{Gaia} DR3}
\label{compgaiaapogee}

To investigate the presence of the `wings' in the comparison between our SED-fitted temperatures and the spectroscopic temperatures from APOGEE (Figure \ref{CompTeff}) we compared the APOGEE temperatures with the other spectroscopic samples used in Section \ref{spectro}. We found a significant number of matches between APOGEE and \textit{Gaia} DR3 ESP-HS (7584 stars) and GSP-Phot (5191 stars) and used these samples to perform this investigation (the overlap between APOGEE and the Houk catalogues was smaller).

\begin{figure*}
\centering
\includegraphics[scale = 0.3]{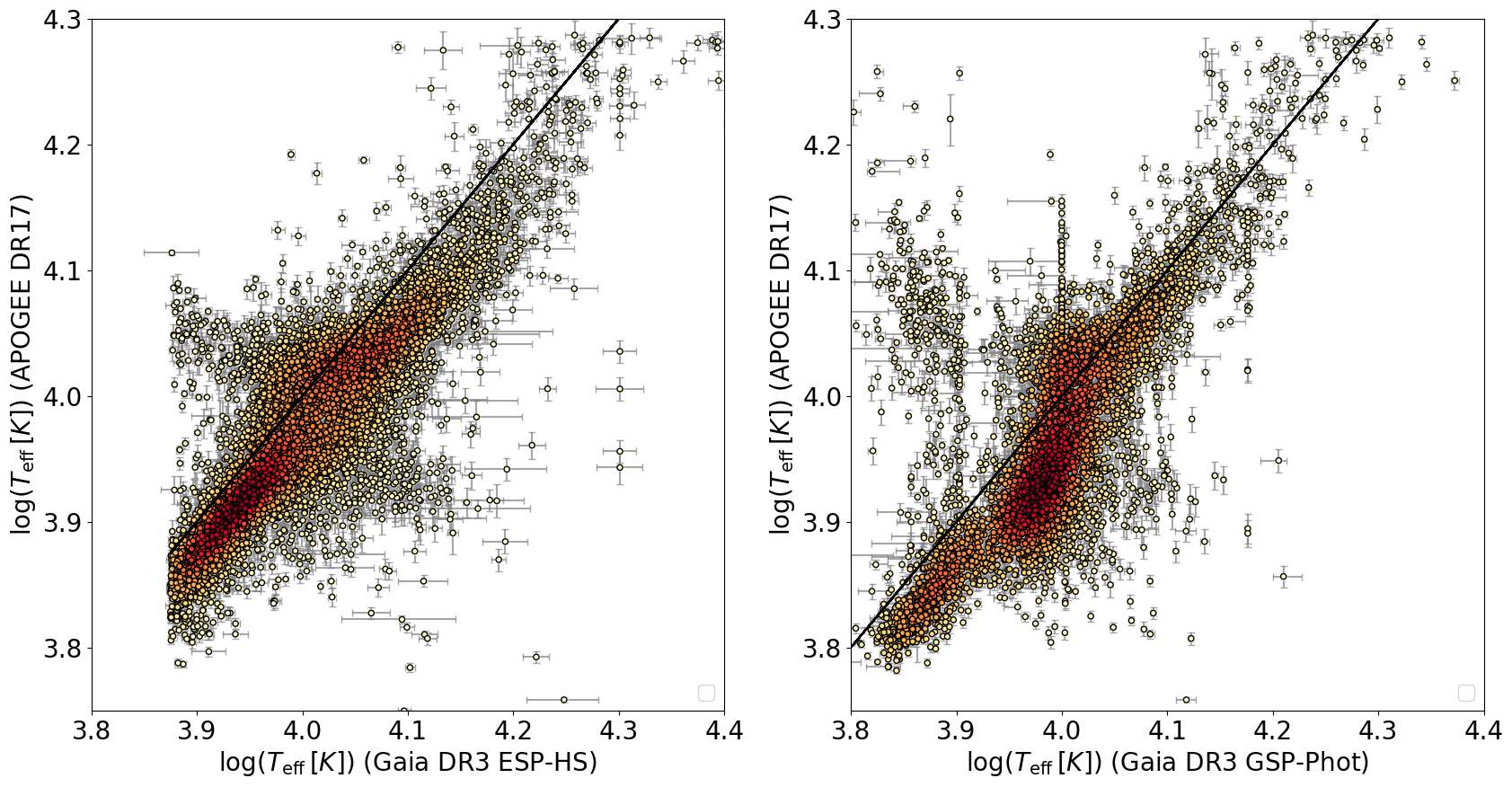}
\caption{Comparison between the spectroscopic temperatures from APOGEE DR17 with those from the \textit{Gaia DR3} ESP-HS (left panel) and GSP-Phot (right panel) modules, with each star colour-coded by its gaussian KDE. A 1:1 line is shown in black. \label{CompGaiaAPOGEE}}
\end{figure*}

Figure \ref{CompGaiaAPOGEE} shows a comparison between the spectroscopic temperatures from APOGEE and \textit{Gaia} DR3. Notably, the `wings' observed in the comparison between APOGEE and our SED-fitted temperatures are also visible in the comparisons between APOGEE and \textit{Gaia} DR3, especially for effective temperatures from ESP-HS. This suggests that the origin of these `wings' is likely to be due to problems with the APOGEE spectral fits. We investigated whether the stars that made up the `wings' had properties that differed in any way to those that made up the bulk of the comparison (e.g., in their projected rotational velocities, a property that could affect the quality of the spectral fits), but no significant difference could be found. Based on this we cannot make any conclusion on the origin of the problem, only that it appears most likely to be due to the APOGEE spectral fits and not due to our SED fits.

We also notice, in Fig. \ref{CompGaiaAPOGEE}, a trend of GSP-Phot to overestimate temperatures for cooler stars and underestimate temperatures for hotter stars, relative to temperatures from APOGEE, a trend that has previously been observed by \citet{Fouesnau2023} and \citet{Avdeeva}. The temperatures from the ESP-HS module, on the other hand, tend to be systematically higher than those from APOGEE.

\section{Comparison between OB stars from our catalogue and the \textit{Gaia} DR3 golden sample}
\label{compgaia}

To further illustrate the differences between our catalogue of OB stars (that we label Q24OB) with the golden sample of OB stars from \textit{Gaia} DR3 \citep{GaiaDR3GoldenSample} (that we label G23), we created four samples as follows:

\begin{itemize}
    \item Sample `Q24OB\_G23' includes the 20,821 SED-fitted OB stars from Q24OB that were found in G23.
    \item Sample `Q24COOL\_G23' includes the 24,398 stars from G23 that are not in Q24OB, but have a successful crossmatch with our catalogue of 195,651 candidate OB stars (i.e., we fit them with $T_{\rm eff} < 10,000$~K).
    \item Sample `G23\_Only' includes the 9768 stars from G23 that have no correspondence at all with any of our catalogues. 
    \item Sample `Q24OB\_Only' includes the 3884 SED-fitted OB stars from Q24OB that were not found in G23. 
\end{itemize}

\begin{figure*}
    \centering
    \includegraphics[scale=0.3]{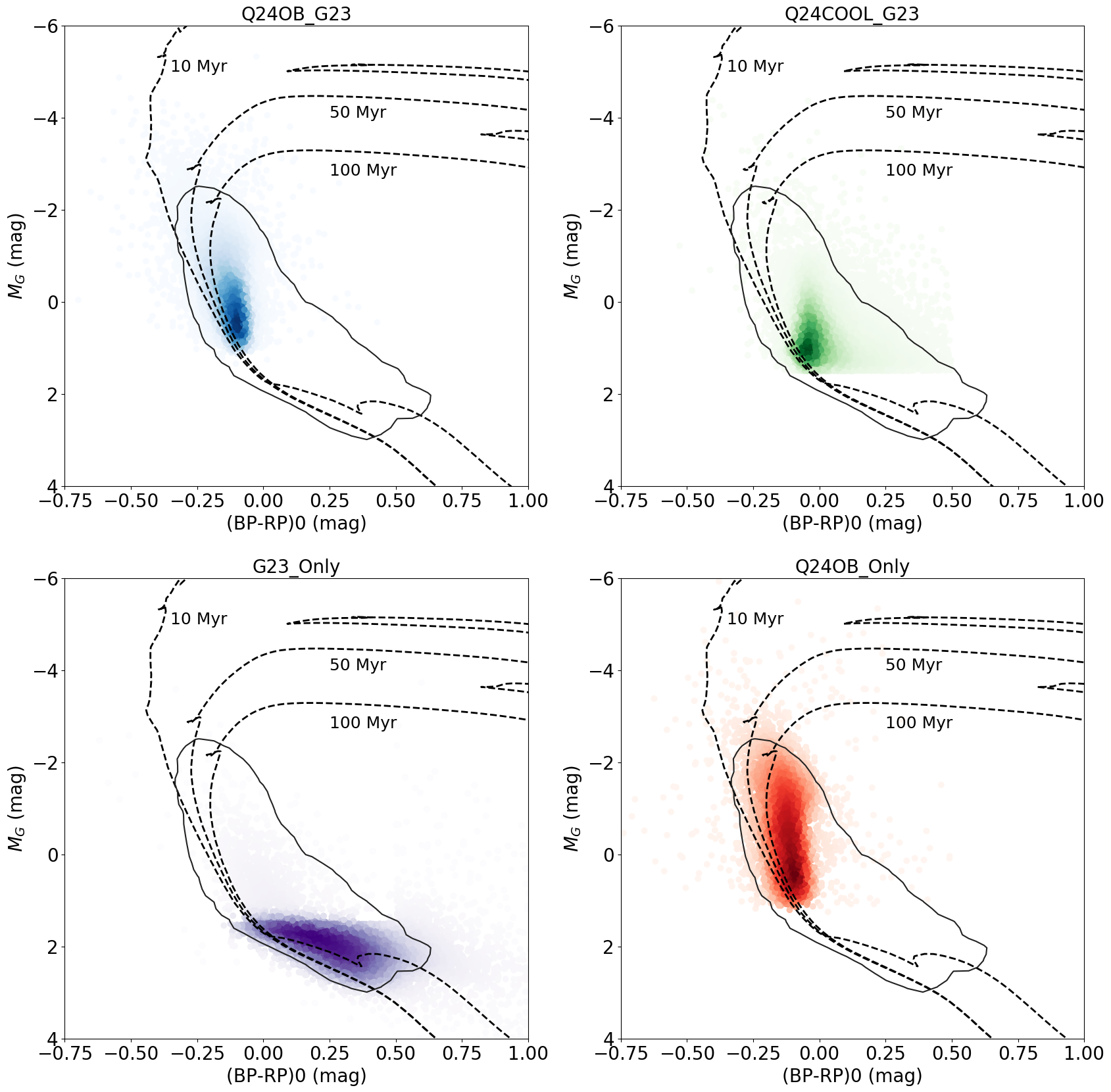}
    \caption{\textit{Gaia} CMDs of the four samples from the comparison with \citet{GaiaDR3GoldenSample}, as outlined in Appendix \ref{compgaia}, colour-coded by their gaussian KDE. In each panel, the dashed lines show 10 Myr, 50 Myr and 100 Myr PARSEC isochrones \citep{Chen2015}. The solid line contour shows the area that encompasses 95\% of all the stars in all four samples combined.}
    \label{CMDGaia}
\end{figure*}

For each sample we have gathered their $M_G$ (from Eq. \eqref{mgsec}) and $(BP-RP)_0$ values as described in Section \ref{maincatalogue}, using the geometric distances from \citet{BailerEDR3} for the G23\_Only subset. This allowed us to construct an unreddened \textit{Gaia} CMD that we have displayed in Fig. \ref{CMDGaia}, and visualise where each subset sit relative to each other in this diagram. 

From Fig. \ref{CMDGaia}, it is clear that the stars in the Q24\_Only sample are genuine OB stars, as their distribution is similar to that of Q24OB\_G23. The distribution of sources in Q24COB\_G23 is offset to lower masses than Q24OB\_G23, supporting our cooler fits to their SEDs. Finally the sources in the G23\_Only sample are unambiguously fainter and redder than any of the other samples, suggesting these are predominantly less massive stars.

% We however point out that both the values of $M_G$ and $(BP-RP)_0$ from Section \ref{maincatalogue} are approximated, and that \citet{GaiaDR3GoldenSample} utilized directly the $(BP-RP)_0$ from the \textit{Gaia} DR3 Apsis modules \citep{Creevey2023}, which also can partly explain these differences.

%%%%%%%%%%%%%%%%%%%%%%%%%%%%%%%%%%%%%%%%%%%%%%%%%%

% Don't change these lines
\bsp	% typesetting comment
\label{lastpage}
\end{document}